\documentclass[twocolumn]{aa}
\usepackage{natbib}
\usepackage{graphicx,amsmath,amssymb,xspace}
\usepackage[svgnames]{xcolor}
\usepackage{subfigure}
\usepackage{caption}
\usepackage{url}
\newcommand{\valpha}{\boldsymbol{\alpha}}
\newcommand{\vrho}{\boldsymbol{\rho}}
\newcommand{\ddroit}{\mathrm{d}}
\newcommand{\wt}[1]{\widetilde{#1}}
\newcommand{\Paren}[1]{\left({#1}\right)}
\newcommand{\Brack}[1]{\left[{#1}\right]}
\newcommand{\Brace}[1]{\left\{{#1}\right\}}
\newcommand{\abs}[1]{\left|{#1}\right|}
\newcommand{\moy}[1]{\left\langle{#1}\right\rangle}

\begin{document}

\title{Simultaneous exoplanet detection and instrument aberration retrieval in multispectral coronagraphic imaging}

\author{M. Ygouf \inst{1} \inst{2} and L. M. Mugnier \inst{1} and D. Mouillet \inst{2} and T. Fusco \inst{1} and J.-L. Beuzit \inst{2}}



\institute{ONERA - The French Aerospace Lab F-92322 Ch\^atillon, France\\
	\and UJF-Grenoble 1 / CNRS-INSU, Institut de Plan\'etologie et d'Astrophysique de Grenoble (IPAG) UMR 5274, Grenoble, F-38041, France}


\date{Received 31 August 2012 / Accepted 16 November 2012}

\abstract {High-contrast imaging for the detection and characterization of exoplanets relies on the instrument's capability to 
block out the light of the host star. Some current post-processing methods for calibrating out the residual speckles use information redundancy offered by multispectral imaging but do not use any prior information on the origin of these speckles.} {We investigate whether 
additional information on the system and 
image formation process can be used to more finely exploit the multispectral information.} {We developed an inversion method in a Bayesian framework  
that is based on an analytical imaging model 
to estimate both the speckles and the object map. The model links the instrumental aberrations to the speckle pattern in the image focal plane, distinguishing between aberrations upstream and downstream of the coronagraph.} {We propose and validate several numerical techniques to handle the difficult minimization 
problems of \emph{phase retrieval} and achieve
a contrast of $10^6$ at 0.2 arcsec from simulated images, in the presence of photon noise.} {This opens up the the possibility of tests on real data where the ultimate performance may override the current techniques if the instrument has good and stable coronagraphic imaging quality. This paves the way for new astrophysical exploitations or even new designs for future instruments.}


\titlerunning{Exoplanet detection and instrument aberration retrieval in multispectral coronagraphic imaging}

\authorrunning{Ygouf et al.} 

\maketitle

\section{Introduction}

Ground-based instruments have now demonstrated the capability 
of detecting planetary mass companions \citep{chauvin-a-04,Lagrange-a-10,Marois-a-08} around bright host stars.
By combining adaptive optics (AO) system and coronagraphs, some first direct detections from the ground have been possible in favorable cases, at large separations and in young systems when low-mass companions are still warm ($\geq$~1000\,K) and therefore not too faint. 
There is a very strong astrophysical case to improve the high-contrast detection capability ($10^5$ for a young giant planet to $10^{10}$ for an earth-like planet in the near infrared) very close to stars ($<$~0.1'' to 1''). 

Several instruments will be capable of performing multispectral imaging and will allow characterizing the planets by measuring their spectra. This is the case of GPI (Gemini)~\citep{Graham-a-07}, Palm 3000 (Palomar)~\citep{Hinkley-a-11}, SCExAO (Subaru)~\citep{Martinache-p-09}, SPHERE (VLT)~\citep{Beuzit-p-08}, and several others that will follow, such as EPICS (E-ELT)~\citep{Kasper-p-08}. 
By combining extreme adaptive optics (Ex-AO) and more accurate coronagraphs than before, the level of star light cancellation is highly improved, leading to a better signal-to-noise ratio.   
Even so, the residual host star light is affected by the instrument aberrations 
and forms a pattern of intensity variations or ``speckle noise'' on the final image. Part of the speckles cannot be calibrated 
because they evolve on various time scales (neither fast enough to smooth down to a halo nor stable enough to remove) and for this reason, these ``quasi-static speckles'' are one of the main limitations for high-contrast imaging.

Several authors have discussed the challenge posed by the elimination of speckle noise in high-contrast multispectral images. 
It can be done by post-processing, after the best possible observations.
Because images are highly spectrally correlated, one can use the wavelength dependence of the speckles to subtract them. 
In the particular case of coronagraphic multispectral imaging, only some empirical methods have been developed to subtract the speckle field from the image in the focal plane. 

We propose an alternative approach based on a parametrized imaging model for the post-processing of multispectral coronagraphic imaging corrected by an extreme 
AO system in the near-infrared domain. The aberrations and bright companions at small separations are estimated jointly in a Bayesian framework.
In particular, it is possible to take advantage of prior information such as a knowledge on the aberration levels.
This 
type of approach will be 
all the more efficient as the instruments improve with lower or more stable aberrations and more efficient coronagraphs.

In section 2, we explain how previous methods 
used the information redundancy 
to suppress the speckles in high-contrast imaging. Then, we describe the advantages of a joint Bayesian estimation of the aberrations in the pupil plane and of the planet map, based on a parametrized model of coronagraphic imaging. 
Section 3 presents the long-exposure coronagraphic imaging model 
that is used to simulate the images and to restore them. 
We also study the case of an approximate model. 
Section 4 describes the proposed Bayesian joint estimation method as well as theoretical and numerical 
problems of an alternating restoration algorithm. In particular, 
we address the strong minimization difficulties associated to the aberration estimation and we propose some solutions. 
In Section 5, our method is validated by restoring images simulated with a perfect coronagraph.

\section{Post-processing speckle subtraction and multispectral imaging}
\begin{figure*}
\begin{center}
\includegraphics[width=4.35cm]{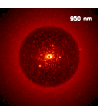}
\includegraphics[width=4.35cm]{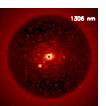}
\includegraphics[width=4.35cm]{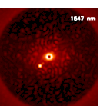}
 \caption{\textbf{Evolution of the speckle field with the wavelength.} Simulated images at 950, 1306, and 1647 \,nm for a $10^3$ stellar flux over planet flux contrast. The dynamic range is adapted to the visualization. The speckle field moves with the wavelength but not with the planet position.}
\label{figure1}
\end{center}
\end{figure*}

Several empirical post-processing methods have already been proposed 
to overcome the problem of detection limitation caused by the 
quasi-static speckles. 
Some of these methods 
used the wavelength dependence of the speckle pattern (Fig.~\ref{figure1}) to estimate it and subtract it from the image,
while preserving both the flux and spectrum of the planet.

\citet{Racine-a-99} suggested to subtract two images at different wavelengths to eliminate both the point-spread function (PSF) and the speckle field in non-coronagraphic images.
The main limitation of this \emph{simultaneous differential imaging} (SDI) method comes from the residuals caused by the evolution of the general PSF profile and of the speckle pattern with wavelength.
These residuals can be reduced by increasing the number of images used for the speckle field subtraction. \citet{Marois-a-00} showed with their \emph{double difference} method that adding another image to the 
SDI theoretically improves the signal-to-noise ratio in the final image of the restored companion. 
The case of multispectral images has been tackled by \citet{Sparks-a-02}, who described the so-called \emph{spectral deconvolution} method in the framework of space-based observations for an instrument combining a coronagraph and an integral-field spectrometer (IFS). The method, 
subsequently 
improved and tested on ground-based non-coronagraphic data by \citet{Thatte-a-07}, is entirely based on a speckle intensity fit by low-order polynomials as a function of wavelength in the focal plane.  
More recently, Crepp et al. (2011) combined this method with the LOCI algorithm, which is based on a linear combination of images~\citep{Lafreniere-a-07b}. They tested this approach to restored on-sky images from the Project 1640 IFS on the Palomar telescope. 
These methods are applicable to any optical system and in particular to those with coronagraphs. However, it is challenging to prevent the planet signals from being eliminated with the speckles 
because the planet presence is not explicitly modeled.

In addition, some information on the measurement system can be very useful to 
distinguish a planet from the speckle field. 
\citet{Burke-a-10} combined classical empirical techniques of differential imaging with a multi-wavelength \emph{phase retrieval} method to estimate the 
aberration pattern in the pupil plane with a simple imaging model without a coronagraph. 
This multi-wavelength \emph{phase retrieval} is nicknamed 
wavelength diversity~\citep{Gonsalves-a-82}. 
Information diversity is obtained here by different wavelengths  whereas it is obtained by introducing a known phase, e.g. defocus, in \emph{phase diversity}. 
But 
in contrast to the \emph{phase diversity}, the 
wavelength diversity does not remove the phase sign ambiguity. 
In \citet{Burke-a-10}, the inversion algorithm is based on a maximum-likelihood estimator,
which measures the discrepancy between the data and an imaging model. 
The minimization of this estimator is all the more difficult as the number of unknowns to estimate 
is high.
This 
problem is overcome by the sparse parametrization of the unknown phases $\phi_{\lambda}$ through the 
optical-path-errors (or aberrations) $\delta$, assuming that the former are achromatic:  $\phi \Paren{\lambda}= 2\pi\delta / \lambda$.
This allows one to exploit jointly the images at all wavelengths to estimate the aberrations efficiently: the map of the unknown optical-path-errors $\delta$ is common to all wavelengths. 
The number of unknowns is thus limited and the problem constrained. 
In the present case, Burke's wavelength diversity method does not apply readily, 
because it assumes non-coronagraphic imaging, whereas 
we consider the highly non-linear case of a coronagraphic imaging model. 

That is why we propose to take advantage of a combined use of wavelength diversity applied in a case of a coronagraphic imaging model, and a Bayesian inversion to jointly estimate the aberrations in the pupil plane and the planet map. 
The joint estimation aims at taking up the challenge of preserving the planets signal. 
An advantage of the Bayesian inversion is that it can potentially include an important regularization diversity to constrain the problem, using for example prior information on the noise, the planet map (position, spectrum, etc.) or the aberrations.
In the Bayesian framework,
the criterion to be minimized is 
the sum of two terms: the data fidelity term, which measures the distance between the data 
and the imaging model, 
and one or more penalty terms.
An important difficulty is to 
define a realistic coronagraphic imaging model, 
that depends on parameters (e.g. aberrations)  that can be either calibrated beforehand or estimated from the data.

\section{Parametric model for multispectral coronagraphic imaging}

To carry out the Bayesian inversion, we need a parametric direct model of coronagraphic imaging. This direct model will also be useful to simulate our test images.

We 
used a non-linear analytical expression of the coronagraphic image as proposed by \citet{Sauvage-a-10}, with an explicit role of the optical aberrations before and after the coronagraph, and turbulence residuals. This model assumes that the coronagraph is ``perfect'' in the sense that the coherent energy is perfectly canceled out. The presence of upstream aberration however, will result in remaining intensity from the star in the image. The aberrations, or optical-path-errors, $\delta$, are assumed to be achromatic as an approximation. The most recent spectro-imagers take 
increasing care to avoid any source of chromatism, 
such as out-of-pupil aberrations, down to a level compatible with contrasts higher than $10^6$. The variable $\alpha \equiv \Paren{\alpha_x,\alpha_y}$  represents the angular position in the focal plane in radians and the variable $\rho \equiv \Paren{\rho_x,\rho_y}$ is the angular position in the pupil plane in radians$^{-1}$. Finally, $ \lambda \rho \equiv \Paren{ \lambda \rho_x, \lambda \rho_y}$ corresponds to a spatial position in the pupil plane in meters.

We recall and discuss this model below.
In particular, we estimate its simplified expression in the asymptotic case of very low phase, with its second-order Taylor expansion. This simplified expression helps to understand the explicit way in which each type of aberration impacts the image. It also helps to identify some important ambiguities with different sets of phases that can produce similar images, 
which will guide the subsequently selected approach for phase retrieval. We also estimate the departure from this low-phase approximation when the phase grows and 
discuss the validity of this approximation in a SPHERE-like case.

\subsection{Imaging model} \label{global_model}
We assume that for an AO-corrected coronagraphic image at the wavelength $\lambda$, the direct model is the following sum of three terms, separating the residual coronagraphic stellar halo, the circumstellar source (for which the impact of coronagraph is neglected),  
and noise $n_\lambda$:
\begin{equation}
i_\lambda\Paren{\alpha} = f_{\lambda}^* \cdot h_\lambda^c\Paren{\alpha} + \Brack{o_\lambda\star h_\lambda^{nc}}\Paren{\alpha} + n_\lambda\Paren{\alpha},
\label{imaging_model}
\end{equation}
where $i_\lambda\Paren{\alpha}$ is the data, i.e., the image to which we have access, $f_{\lambda}^*$ is the star flux at wavelength lambda and $h_\lambda^{nc}\Paren{\alpha}$, the non-coronagraphic 
PSF, which can be estimated separately. 
Solving the inverse problem is finding the unknowns, namely the object $o_\lambda\Paren{\alpha}$ and the speckle field $h_\lambda^c\Paren{\alpha}$, which we also call the ``coronagraphic PSF''. 

\subsection{Long-exposure coronagraphic PSF model} \label{model_existant}
\begin{figure*}
  \begin{center}
 
\begin{picture}(10,80)(200,80)%
\includegraphics[width=14cm]{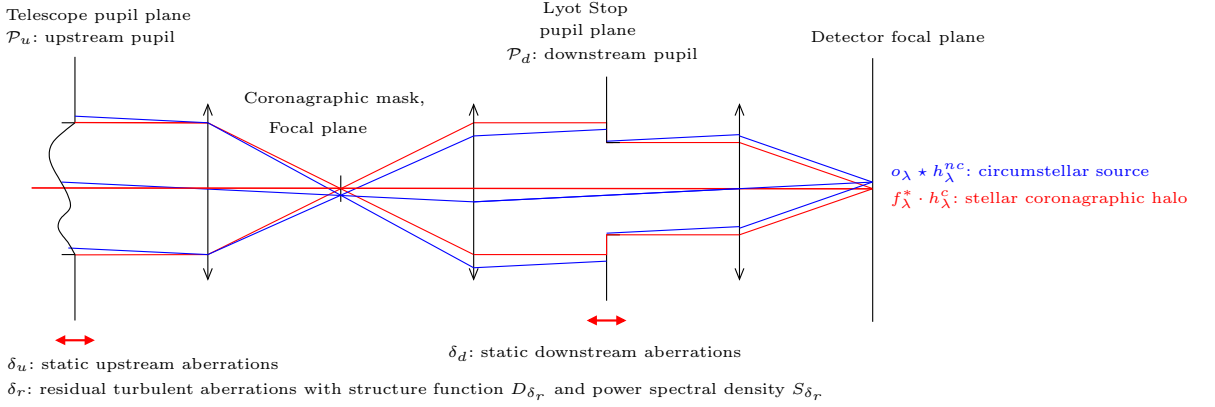}%
\end{picture}%

\setlength{\unitlength}{4144sp}%
\begingroup\makeatletter\ifx\SetFigFont\undefined%
\gdef\SetFigFont#1#2#3#4#5{%
  \reset@font\fontsize{#1}{#2pt}%
  \fontfamily{#3}\fontseries{#4}\fontshape{#5}%
  \selectfont}%
\fi\endgroup%

\begin{picture}(6901,-2881)(3601,-2575)
\put(6950,-1461){\makebox(0,0)[lb]{\smash{{\SetFigFont{6}{8.4}{\rmdefault}{\mddefault}{\updefault}pupil plane}}}}
\put(9030,-2461){\makebox(0,0)[lb]{\smash{{\SetFigFont{6}{8.4}{\rmdefault}{\mddefault}{\updefault}{\color[rgb]{1,0,0}$f_{\lambda}^* \cdot h_{\lambda}^{c}$: stellar coronagraphic halo}%
}}}}
\put(9030,-2300){\makebox(0,0)[lb]{\smash{{\SetFigFont{6}{8.4}{\rmdefault}{\mddefault}{\updefault}{\color[rgb]{0,0,1}$o_{\lambda} \star h_{\lambda}^{nc}$: circumstellar source}%
}}}}
\put(3791,-1371){\makebox(0,0)[lb]{\smash{{\SetFigFont{6}{8.4}{\rmdefault}{\mddefault}{\updefault}{\color[rgb]{0,0,0}Telescope pupil plane}%
}}}}
\put(3791,-1500){\makebox(0,0)[lb]{\smash{{\SetFigFont{6}{8.4}{\rmdefault}{\mddefault}{\updefault}{\color[rgb]{0,0,0}$\mathcal{P}_u$: upstream pupil}%
}}}}
\put(5201,-1866){\makebox(0,0)[lb]{\smash{{\SetFigFont{6}{8.4}{\rmdefault}{\mddefault}{\updefault}{\color[rgb]{0,0,0}Coronagraphic mask, }%
}}}}
\put(5350,-2046){\makebox(0,0)[lb]{\smash{{\SetFigFont{6}{8.4}{\rmdefault}{\mddefault}{\updefault}{\color[rgb]{0,0,0}Focal plane}%
}}}}
\put(6971,-1326){\makebox(0,0)[lb]{\smash{{\SetFigFont{6}{8.4}{\rmdefault}{\mddefault}{\updefault}{\color[rgb]{0,0,0}Lyot Stop}%
}}}}
\put(6756,-1600){\makebox(0,0)[lb]{\smash{{\SetFigFont{6}{8.4}{\rmdefault}{\mddefault}{\updefault}{\color[rgb]{0,0,0}$\mathcal{P}_d$: downstream pupil}%
}}}}
\put(8556,-1500){\makebox(0,0)[lb]{\smash{{\SetFigFont{6}{8.4}{\rmdefault}{\mddefault}{\updefault}{\color[rgb]{0,0,0}Detector focal plane}%
}}}}
\put(6411,-3380){\makebox(0,0)[lb]{\smash{{\SetFigFont{6}{8.4}{\rmdefault}{\mddefault}{\updefault}{\color[rgb]{0,0,0}$\delta_d$: static downstream aberrations}%
}}}}
\put(3801,-3450){\makebox(0,0)[lb]{\smash{{\SetFigFont{6}{8.4}{\rmdefault}{\mddefault}{\updefault}{\color[rgb]{0,0,0}$\delta_u$: static upstream aberrations}%
}}}}
\put(3801,-3600){\makebox(0,0)[lb]{\smash{{\SetFigFont{6}{8.4}{\rmdefault}{\mddefault}{\updefault}{\color[rgb]{0,0,0}$\delta_r$: residual turbulent aberrations with structure function $D_{\delta_r}$ and power spectral density $S_{\delta_r}$}%
}}}}
\end{picture}%

\bigskip
\bigskip
\bigskip
\bigskip
\bigskip
\smallskip

    \caption{\textbf{Optical scheme of a coronagraphic imager.} The upstream and
      downstream static aberrations, and the adopted notations are
      denoted~$\delta_u$ and~$\delta_d$.~$\mathcal{A}_i(\valpha)$
      denote focal plane complex amplitudes, whereas~$\Psi_i(\vrho)$ denotes
      pupil plane amplitudes.}
    \label{fig-coro_ab}
  \end{center}
\end{figure*}

A model description of $h_{\lambda}^c\Paren{\alpha}$ directly depends on the turbulence residuals and optical wave-front errors. 
After previous works to model non-coronagraphic PSFs~\citep{perrin-a-2003} and coronagraphic PSFs~\citep{Cavarroc-a-06,Soummer-a-07}, \citet{Sauvage-a-10} proposed an analytical expression for the coronagraphic image with a distinction between upstream and downstream aberrations (cf. Equation~(\ref{model_jeff}) in Appendix B). 
The considered optical system is composed of a telescope, a perfect coronagraph, and a detector plane (cf. Figure~(\ref{fig-coro_ab})). Some residual turbulent aberrations $\delta_r(\rho,t)$ are introduced in the telescope pupil plane. $\delta_r(\rho,t)$ is assumed to be temporally zero-mean, stationary and ergodic. Because we only consider exposure times 
that are long with respect to turbulence timescales, these turbulent aberrations contribute only through their statistical spatial properties: power spectral density $S_{\delta_r}(\alpha)$ or structure function $D_{\phi_r}$. The  static aberrations are separated into two contributions: the aberrations upstream of the coronagraph $\delta_u(\rho)$, in the telescope pupil plane $\mathcal{P}_u(\rho)$ and the aberrations downstream of the coronagraph $\delta_d(\rho)$ in the Lyot Stop pupil plane $\mathcal{P}_d(\rho)$. 
The perfect coronagraph is defined as an optical device that subtracts a centered Airy pattern of maximal energy from the electromagnetic field.
Finally, the 
coronagraphic PSF depends on three parameters 
that define our system: the aberration maps $\delta_u$, $\delta_d$ and, the residual phase structure function $D_{\phi_r}$.


\subsubsection{Approximate long-exposure coronagraphic model in the low-phase regime} \label{sss:approximate_model}
Because the analytical expression for $h_\lambda^c$ is a highly non-linear function of the aberrations~\citep{Sauvage-a-10}, we derived and studied the relevance of an approximate expression for this model~\citep{Ygouf-p-10a}. Approximate coronagraphic imaging models have been derived in several works.
\citet{Cavarroc-a-06} have developed a short-exposure expression and showed by simulations that the main limitation comes from the static aberrations and particularly the aberrations upstream of the coronagraph. Here, we consider a long-exposure imaging model and confirm analytically the dominance of the upstream aberrations.
\citet{Soummer-a-07} have developed a two-term expression with one static 
and one turbulent term. 
Nevertheless, these terms are not explicitly linked to the aberrations, which is what we are interested in.

Assuming that all 
phases are small and that the spatial 
means of $\phi_u(\rho)$ and $\phi_d(\rho)$ are equal to zero on the aperture, we derive a second-order Taylor expansion of expression 24 of \citet{Sauvage-a-10}:
\begin{align}\label{eq:approximate_model}
\Brack{h_{\lambda}^c}^{app} \Paren{\alpha} 
&= \Paren{\frac{2\pi}{\lambda}}^2 \Brace{ \abs{\wt{\mathcal{P}_d \Paren{\lambda \rho}} \star \wt{\delta_u \Paren{\lambda \rho} }}^2 } \nonumber\\
&+   \Paren{\frac{2\pi}{\lambda}}^2 \Brace{\abs{\wt{\mathcal{P}_d\Paren{\lambda \rho}}}^2 \star S_{\delta_r} \Paren{\alpha} } \nonumber\\
&-  \Paren{\frac{2\pi}{\lambda}}^2 \Brace{\moy{\abs{P\Brack{\delta_r \Paren{\lambda \rho,t}}}^2}_t \cdot \abs{\wt{\mathcal{P}_d\Paren{\lambda \rho}}}^2 } \nonumber\\
&+  o\Paren{\delta^2},
\end{align}
where \small $\wt{\mathcal{P}_d \Paren{\lambda \rho}}$ \normalsize and \small $\wt{\delta_u \Paren{\lambda \rho} }$ \normalsize are the Fourier transforms of the downstream pupil and upstream aberrations 
and \small $P\Brack{\delta_r \Paren{\lambda \rho,t}}$ denotes the piston of the aberration map \small $\delta_r \Paren{\lambda \rho,t}$\normalsize.  \small $\Brace{\moy{\abs{P\Brack{\delta_r \Paren{\lambda \rho,t}}}^2}_t \cdot \abs{\wt{\mathcal{P}_d\Paren{\lambda \rho}}}^2} $ \normalsize is a corrective term that compensates for the fact that \small $\delta_r \Paren{\lambda \rho,t}$ \normalsize is stationary and thus non-piston-free on the aperture at every instant. 
\small $\abs{\wt{\mathcal{P}_d\Paren{\lambda \rho}}}^2$ \normalsize is the Airy pattern formed by the pupil \small $\mathcal{P}_d\Paren{\lambda \rho}$ \normalsize.\\

This approximate expression brings physical insight 
into the long-exposure 
coronagraphic PSF model of Sauvage et al.:
\begin{itemize}
\item The speckle pattern scales radially in $\lambda$ within the approximate model and evolves as $1 / \lambda^2$ in intensity in the data cube. It is consistent with the analysis of \citet{Sparks-a-02}, who 
performed fits of low-order polynomials as a function of the wavelength after rescaling radially.
\item The approximate expression can be separated into one static 
and one turbulent term. This is consistent with the analysis of \citet{Soummer-a-07} with the advantage that these terms depend on the parameters of interest. 
The turbulent term is simply the turbulent aberration power spectral density, as seen at the resolution of the instrument, i.e., convolved by the output pupil Airy pattern. The static term is explicitly a function of the upstream aberrations.
\item The downstream aberrations do not appear in the static term. This confirms that the role of the aberrations upstream and downstream of the coronagraph is very different and that upstream aberrations are dominant in the final image. 
\item Four equivalent upstream aberration sets, $\delta_u(\rho)$, $\delta_u(-\rho)$, $-\delta_u(\rho)$ and $-\delta_u(-\rho)$, which we call 
quasi-equivalent aberration maps in the following, lead to the same image (cf. Appendix A).
This item is 
discussed in more detail in section~\ref{sss:4phases}.
\item By using this approximate expression for $h_\lambda^c$ in the imaging model~(\ref{imaging_model}), we can see that there is a degeneracy between the value of the star flux and the rms value of the aberration map, if there is no turbulence.
Indeed, without turbulent aberrations, the approximate model multiplied by the star flux can be written as 
\begin{align} \label{indetermination}
 f_\lambda^* \cdot h_\lambda^c(\delta_u) =  f_\lambda^* \cdot \frac{\Paren{2\pi}^2}{\lambda^2}  \cdot \abs{\wt{\mathcal{P}}_d \star \wt{\delta}_u}^2.
\end{align}
This is 
discussed in section~\ref{sss:starting_point} in greater depth.
\end{itemize}

\subsubsection{Discussion} \label{sss:discussion}
Because of the complexity of the long-exposure 
coronagraphic PSF model of Sauvage et al., we first 
considered using the approximate model in our inversion algorithm to decrease the number of unknowns to estimate and 
simplify the criterion to minimize. But a study of this approximate model showed that the resulting image is too different from the one simulated with the Sauvage et al. expression: 
 computing the root mean square of the difference between the two images leads to 
a substantial error of typically 30\% in SPHERE-like conditions~\citep{Ygouf-p-10a}. 

\begin{figure*}[htbp]
\begin{center}
\includegraphics[trim= 0mm 0mm 0mm 4.5mm, clip, width=10cm]{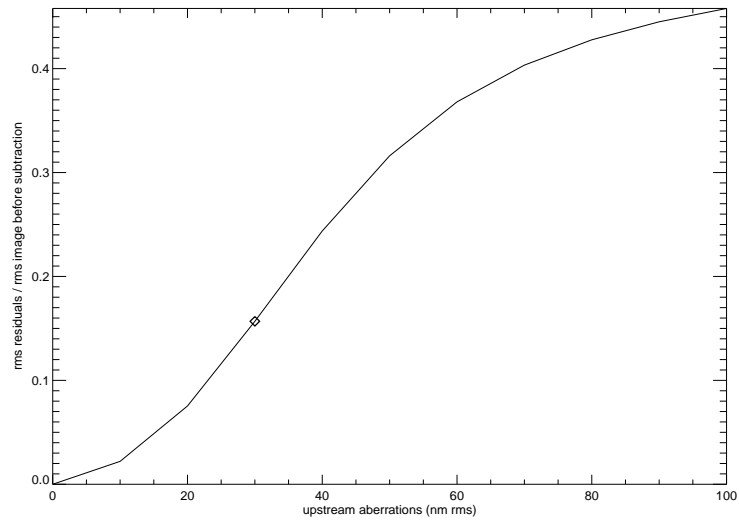}
\caption{\textbf{Speckle non-centrosymmetry}. Evolution of the level of residuals 
with respect to the rms value of the upstream quasi-static aberrations in an image simulated with the model of Sauvage et al. These residuals correspond to the antisymmetric part of the image.}
\label{f:evolution_part_speckles_centro}
\end{center}
\end{figure*}

\begin{figure*}
\begin{center}
\includegraphics[width=13cm]{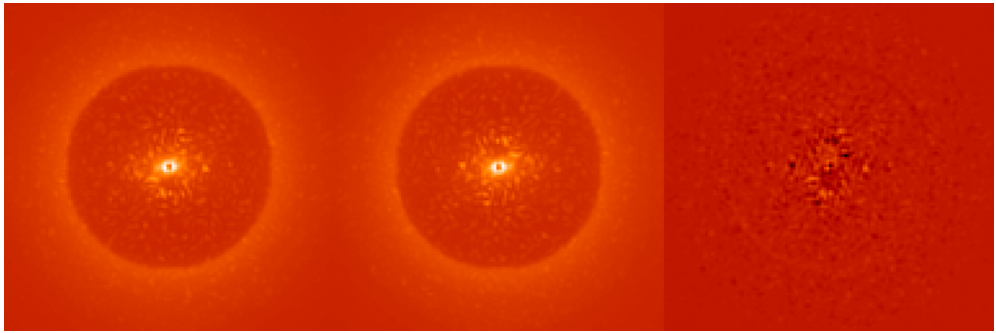}
\caption{\textbf{Speckle non-centrosymmetry}. In the same dynamic range: (left) simulated image, (center) 180 degree rotated version of the image and (right) residuals after combination of the image with the 180 degree rotated version of itself. The gain on the rms value after the subtraction is about 6.}
\label{f:centrosymmetry}
\end{center}
\end{figure*}

\paragraph{Speckle non-centrosymmetry.}
A 
substantial part of this error 
arises because in the approximate model, the quasi-static speckles are purely centrosymmetric. 
Thus, if we eliminate the centrosymmetric part of the image by combining it with a 180 
degree rotated version of itself as follows: 
\begin{align}
i_{\text{antisym}} = \frac{i - i_{180°}}{2},
\end{align}
there are no residuals with this model, i.e. $i_{\text{antisym}}=0$. 
But 
this is not the case with a more physically realistic image such as the one simulated with the model of Sauvage et al.. 
The level of residuals after such a subtraction is determined by the quantity of upstream aberrations, as we can see 
in Figure~\ref{f:evolution_part_speckles_centro}.
For example, with 30\,nm rms of upstream aberrations, the level of residuals is about six times lower than the rms value of the simulated image (cf. Figure~\ref{f:centrosymmetry}).
Thus, for a 
significant quantity of upstream aberrations, using the model of Sauvage et al. rather than the approximate model for an inversion, should lead to 
fewer residuals on the final image. 

\begin{figure*}
\begin{center}
\includegraphics[width=13cm]{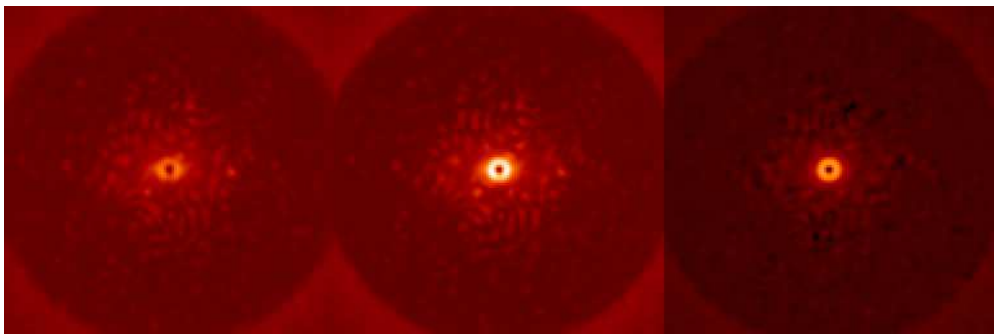}
\caption{\textbf{Speckle dilation}. In the same dynamic range: (left) image at 950\,nm rescaled at 1650\,nm and multiplied by the $\gamma$ coefficient, image at 1650\,nm (center) before and (right) after speckle subtraction. The gain on the rms value after the subtraction is 2.5.}
\label{f:speckles_dilation}
\end{center}
\end{figure*}

\paragraph{Speckle dilation.}
Another difference between the two models tips the balance 
toward the model of Sauvage et al.. Indeed, in the approximate model the speckle dilation is powered by the $1/ \lambda^2$ factor.
If we subtract an image at 950\,nm 
from its 1650\,nm-dilation, there are no residuals with this model. The same operation with the model of Sauvage et al. (cf. Figure~\ref{f:speckles_dilation}) 
shows some residuals, at a level 2.5 lower than the rms value of the simulated image, which attests that it is not a pure speckle dilation.
To do this, we 
took the images at the minimum and maximum wavelengths and 
rescaled the image at 950\,nm with respect to the image at 1650\,nm. Finally, we 
performed the following spectral differences between the two images:
$i_{\text{diff}_{1650}} = i_{1650\,\text{nm}} - \gamma i_{950\,\text{nm}}$, 
where $\gamma$ is a coefficient 
that minimizes the squared difference $\abs{i_{\text{max}}- \gamma i_{{\text{min}}}}^2$, 
and is given by~\citet{Cornia-p-10}: 
\begin{align}
\gamma = \frac{ \sum_{\rho} i_{950\,\text{nm}}(\rho)i_{1650\,\text{nm}}(\rho) } { \sum_{\rho} i_{1650\,\text{nm}}^2(\rho) }.
\end{align}

A fine model of the speckle field must be able to account for deviations form centrosymmetry and for deviations from a radial scaling of the speckles proportional to the wavelength. That is why we 
adopted the model of Sauvage et al. rather than the approximate one in the inversion.

\subsubsection{Assumptions on the long-exposure coronagraphic PSF model} \label{sss:model_simplification}
The information we 
obtained from the approximate model study 
helped us define some key assumptions for the success of the speckle field estimation with the Sauvage et al. long-exposure coronagraphic PSF model.

Because they have 
quite a different impact on the final image, it is important to distinguish the aberrations upstream and downstream of the coronagraph. 
The effect of the downstream aberrations is lower than that of the upstream aberrations, and furthermore, in 
predicted systems such as SPHERE, they are expected to be much more stable and easier to calibrate than upstream aberrations. 
Additionally, because we consider long-exposure images, the residual turbulent aberrations will be averaged to form a smooth halo that is easily distinguishable from a planet. Furthermore, the statistical quantity  $D_{\phi_r}$, which characterizes this halo, will be measured through the 
AO system wavefront sensor \citep{Veran-a-97}.
Therefore, we here 
assumed that both the static downstream aberrations and the residual turbulent aberrations are calibrated and known. This decreases the number of unknowns 
because the only aberration map to estimate in order to 
access 
the 
coronagraphic PSF is the quasi-static upstream aberrations. 
We 
therefore denote the long-exposure 
coronagraphic PSF by $h_\lambda^c\Paren{\delta_u ; \delta_d, D_{\phi_r}}$ instead of $h_\lambda^c\Paren{\delta_u ,\delta_d, D_{\phi_r}}$ to underline the fact that $\delta_d$ and $D_{\phi_r}$ are assumed to be known.

An advantage of our approach is that these assumptions can evolve. The formalism will allow us to refine our method if we finally decide to estimate either the downstream aberrations or the residual turbulent aberrations. Thus, we can 
slowly increase the complexity of the problem in anticipation of using real data from SPHERE or from another instrument.


\section{Joint estimation of wavefront and object algorithm and minimization strategy}\label{s:minimization_issues}
This section introduces the criterion to be minimized (\ref{criterion}) 
as well as 
the regularization elements 
that were used to constrain the problem for the present validations (\ref{regularisations}). 
The minimization algorithm is then described, stressing the two stages 
that constitute its core (\ref{iterative_algorithm}). 
One of these stages presents some convergence difficulties. A minimization strategy is described in Subsection (\ref{ss:phase_retrieval}).
The chosen optimizer in described in Subsection~\ref{VMLM}.

\subsection{Definition of the criterion to be minimized 
}\label{criterion}
Following the Bayesian inverse problem approach, solving the inverse problem consists 
of finding the unknowns, firstly the object characteristics $o\Paren{\alpha,\lambda} =  \Brace{o_\lambda\Paren{\alpha}}_{\lambda}$, secondly the parameters of the speckle field $h_\lambda^c(\delta_u; \delta_d, D_{\phi_r})$ and $f^*\Paren{\lambda} =  \Brace{f_{\lambda}^*}_{\lambda}$, which are the most likely given the data and our prior information about the unknowns.
This 
can be reduced to minimizing the following criterion:
\begin{align} \label{critere}
 J(o, f^*, \delta_u) &= \sum_\lambda \sum_{\alpha} \frac{1}{2\sigma_{n,\lambda}^2\Paren{\alpha} }|i_\lambda - f_{\lambda}^* \cdot h_\lambda^c(\delta_u; \delta_d, D_{\phi_r})  \nonumber\\
&- o_\lambda \star h_\lambda^{nc}(\delta_u; \delta_d, D_{\phi_r})|^2\Paren{\alpha} \nonumber\\
&+ R_{o} + R_{f^*} + R_{\delta}+ \cdot \cdot \cdot .
 \end{align}
This criterion is the sum of two terms: the data fidelity term, which measures the distance between the data and the imaging model, 
and a non-exhaustive list of regularization terms on our unknowns $R_{o}$, $R_{f^*} $, $R_{\delta}$. 
We consider that the noise is the sum of a photonic contribution and a detector contribution, that it is white and approximately Gaussian, which is a valid approximation for the flux as considered in this application. 
The noise variance is assumed to be known here and if it were not, it could be estimated as $\hat{\sigma}_{n,\lambda}^2 = \hat{\sigma}_{ph,\lambda}^2 + \hat{\sigma}_{det,\lambda}^2 $~\citep{mugnier-a-04}, where $\hat{\sigma}_{ph,\lambda}^2 = \max(i_\lambda, 0)$ is the photon noise variance and $\hat{\sigma}_{det,\lambda}^2$ is the detector noise variance previously calibrated. It is assumed that the noise is not correlated from pixel to pixel or between images.

The star flux at each wavelength can be analytically estimated from the criterion provided the regularization on flux is quadratic or absent. In the latter case, the maximum likelihood solution being given by $\frac{\partial{J}}{\partial{f_\lambda^*}} = 0$, we 
obtain: 
 \begin{align} \label{flux_analytique}
\hat{f}_\lambda^*\Paren{o_{\lambda}, \delta_u} = \frac{ \sum_{\alpha}{ \Brack{h_\lambda^c \Paren{i_\lambda - o_\lambda \star h_\lambda^{nc}}/\sigma_{n,\lambda}^2}} \Paren{\alpha}}{\sum_{\alpha} \Brack{ {\Paren{h_\lambda^c}^2} / \sigma_{n,\lambda}^2 } \Paren{\alpha}}.  
 \end{align}
Thanks to this analytical expression, the criterion to be minimized is that of Eq.~(\ref{critere}) with $f_{\lambda}^*$ replaced by $\hat{f}_\lambda^*$, which will be denoted by $J'\Paren{o,\delta_u}$ and depends explicitly on $o_{\lambda}$ and $\delta_u$ only. 


%

Estimating both the object and the aberrations from a single image is a highly underdetermined problem. 
Any diffraction-sized feature in the halo can be interpreted either as a circumstellar point-source or as a part of the stellar speckle halo. \color{black}
This ambiguity can be decreased by using multispectral images but it is not sufficient.  
It is thus necessary to regularize the problem by adding more constraints.
This is the role of the regularization terms. 
In this paper, we study the case of constraints on the object 
that is sufficient 
for the simulated images we used, but we should keep in mind that it is also possible to use constraints on the aberration map. 

%
%

\subsection{Regularization terms and constraints}\label{regularisations}

We describe 
below the regularization terms 
that we used for the validation tests of this paper. 
Other regularizations could be chosen depending on the kind of images to be processed.

\subsubsection{Regularization on the object $R_{o}$}\label{regul L1-L2}

A regularization on the object is fundamental to help compensating for the degeneracy 
that exists in the inversion between the aberrations and the object.
By penalizing the energy in the object map, it favors the energy in the aberration map and prevents the speckles from being mistaken for a planet.

The regularization term $R_{o}$ includes the prior spatial and spectral information we have on the object. 
We chose here an L1-L2 white spatial regularization, which assumes independence between the pixels~\citep{meimon-a-09} because we are mainly looking for point sources. 
We 
used an L1-L2 regularization rather than a true L1 regularization to keep a differentiable criterion, which simplifies the minimization problem.
The spectral prior is based on smoothness of the object spectrum. We currently assume that the object is white (constant spectrum) but as the final aim is to extract some spectra, for future validations we will use a L2 correlated spectral regularization~\citep{thiebaut-p-06}, which will involve the differences between the spectrum at neighboring wavelength at each pixel and will enforce smoothness on the object spectrum.

The regularization on the object is necessary to 
obtain a sparse object. Without the regularization 
many residuals 
remain on the estimated object.

\subsubsection{Positivity and support constraints on the object}\label{contrainte_positivite}

The object intensity map is a set of positive values, which is an important prior information.
One should therefore enforce a positivity constraint on the object.	
This constraint can be implemented in various ways, such as criterion minimization under the positivity constraint, reparameterization of the object, or explicit modification of the a priori probability distribution (e.g., addition of an entropic term).
\citet{mugnier-a-04} have found that the best way to ensure positivity, with respect both to speed and to not introducing local minima, is to directly minimize the criterion under this constraint. We 
proceeded similarly.

Because the star light is concentrated around the optical axis, the flux is 
essentially estimated on this very bright region (cf. Equation~\ref{flux_analytique}). 
But if there are too many residuals on the object, the flux estimation can be biased.
Thus, 
imposing, as is physically meaningful, 
that the object is null very close to the star, in a region of typically $3 \lambda /D$ radius, helps us estimate the star flux accurately.


\subsubsection{Regularization on the star flux $R_{f^*}$}\label{regul_flux}

To make the minimization more robust, it can be useful to constrain the flux estimation to physical values. Indeed, if there are no turbulent aberrations, there is an indetermination between the star flux value and the phase rms value in the approximate model of Eq. (\ref{eq:approximate_model}). Consequently, for aberrations with very 
low rms values, the 
coronagraphic PSF $h_{\lambda}^c$ is close to zero and thus the analytical flux estimate given by Eq. (\ref{critere}) can diverge. The presence of known turbulent aberrations naturally constrains the flux value, but to make the method more robust, we chose to prevent the flux from diverging. In practice, we 
regularized the flux estimation by the following quadratic metric: $R_{f^*}= \frac{\Paren{f_\lambda^* - f_0}^2} {2 \sigma_{f,\lambda}^2}$. This metric can be interpreted as a Gaussian prior low on the flux, 
but its role is not as essential for the criterion minimization as that of the L1-L2 regularization and the positivity constraint.
This leads to the following expression for the analytic star flux:
 \begin{align}\label{flux_regularise}
\hat{f}_\lambda^* = \frac{\sum_{\alpha} h_\lambda^c  \Paren{i_\lambda - o_\lambda \star h_\lambda^{nc}}/ \sigma_{n,\lambda}^2 + f_0 / \sigma_{f,\lambda}^2 }
   {\sum_{\alpha} \Paren{h_\lambda^c}^2 / \sigma_{n,\lambda}^2 + 1/ \sigma_{f,\lambda}^2}.
 \end{align}
In practice, we 
chose a very 
high standard deviation $\sigma_{f,\lambda}= 100 \times \sum_{\alpha}  i_{\lambda}$, 
to avoid biasing the flux. With 
this standard deviation, we can choose any mean flux, for example, $f_0 = 0$. This is sufficient to avoid the division by zero in the flux computation and thus the flux divergence. 


\subsection{Iterative algorithm}\label{iterative_algorithm}
\color{black}
The structure of the joint criterion of Eq.~(\ref{critere}) prompted us to adopt an estimation of wavefront and object 
that alternates between estimation of the aberrations, assuming that the object is known (multispectral \emph{phase retrieval}) and estimation of the object assuming that the aberrations are known (\emph{multispectral deconvolution}). 
 

\paragraph{Multispectral deconvolution.}
For given aberrations, we define the following intermediate data where the (assumed known) stellar halo is subtracted, keeping then only the circumstellar object as seen in classical imaging: $i''_\lambda  = i_\lambda - f_{\lambda}^* \cdot h_\lambda^c(\delta_u; \delta_d, D_{\phi_r}) $.
By inserting 
these intermediate data into the the criterion of Eq.~(\ref{critere}), we 
obtain: 
 \begin{align} \label{critere1}
J''(o, f^*, \delta_u) &= \sum_\lambda \sum_{\alpha} \frac{1}{2\sigma_{n,\lambda}^2\Paren{\alpha} }|i''_\lambda 
- o_\lambda \star h_\lambda^{nc}(\delta_u; \delta_d, D_{\phi_r})|^2\Paren{\alpha} \nonumber\\
&+ R_{o} + R_{f^*} + R_{\delta}+ \cdot \cdot \cdot ,
 \end{align}
which shows that the problem at hand is a \emph{non-myopic multispectral deconvolution} of images $i''_\lambda$.
The chosen regularization leads to a convex criterion~\citep{mugnier-a-04} and thus to a unique solution for a given set of aberrations and a given object regularization.  

\paragraph{Phase retrieval.}
If we replace the 
intermediate data 
$i'_\lambda  = i_\lambda - o_\lambda \star h_\lambda^{nc}(\delta_u; \delta_d, D_{\phi_r})$ 
into the the criterion of Eq.~(\ref{critere}), we 
obtain: 
\begin{align} \label{critere2}
J'(o, f^*, \delta_u) &= \sum_\lambda \sum_{\alpha} \frac{1}{2\sigma_{n,\lambda}^2\Paren{\alpha} }|i'_\lambda 
- f_{\lambda}^* \cdot h_\lambda^c(\delta_u; \delta_d, D_{\phi_r})  |^2\Paren{\alpha} \nonumber\\
& + R_{o} + R_{f^*} + R_{\delta}+ \cdot \cdot \cdot ,
 \end{align}
which shows that the problem at hand is essentially a \emph{phase retrieval} problem.
In this \emph{phase retrieval} stage, 
the combination of a high number of parameters to estimate (typically $10^3$, see Section~\ref{s:validations}) and of a highly non-convex criterion
complicates the problem. 
To 
avoid local minima, several numerical solutions resulting from a fine understanding of the imaging process are necessary and 
are described below.

\color{black}
 
\subsection{Phase retrieval: dealing with local minima}\label{ss:phase_retrieval}


 
\subsubsection{Choice of an appropriate starting point: very small random phase}\label{sss:starting_point}
To keep the computation time reasonable, we 
used the local descent algorithm described in Subsection~\ref{VMLM} to minimize the criterion. Because the latter is highly non-convex, the chosen starting point 
so that we are fully in the conditions where the Taylor expansion developed in  \ref{sss:approximate_model} is valid and where the criterion is less non-convex. 
It allows the algorithm to avoid many wrong directions, and thus many local minima.  
As the algorithm converges, the upstream aberration rms value increases 
toward its true value and a gradual non-linearity of the model is 
gradually introduced.

\begin{figure*}
\begin{center}

\subfigure[]{

\includegraphics[width=13.3cm]{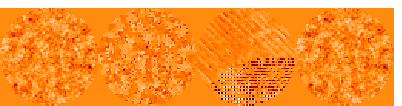}
\setlength{\unitlength}{4144sp}%
\begingroup\makeatletter\ifx\SetFigFont\undefined%
\gdef\SetFigFont#1#2#3#4#5{%
  \reset@font\fontsize{#1}{#2pt}%
  \fontfamily{#3}\fontseries{#4}\fontshape{#5}%
  \selectfont}%
\fi\endgroup%

\begin{picture}(0,0)
\put(-5950,1300){\makebox(0,0)[lb]{\smash{{\SetFigFont{7}{8.4}{\rmdefault}{\mddefault}{\updefault}``true''}}}}
\put(-4500,1300){\makebox(0,0)[lb]{\smash{{\SetFigFont{7}{8.4}{\rmdefault}{\mddefault}{\updefault}(a)}}}}
\put(-3000,1300){\makebox(0,0)[lb]{\smash{{\SetFigFont{7}{8.4}{\rmdefault}{\mddefault}{\updefault}(b)}}}}
\put(-1500,1300){\makebox(0,0)[lb]{\smash{{\SetFigFont{7}{8.4}{\rmdefault}{\mddefault}{\updefault}(c)}}}}
\put(-5800,-150){\makebox(0,0)[lb]{\smash{{\SetFigFont{7}{8.4}{\rmdefault}{\mddefault}{\updefault} $\text{rms}^{\text{``true''}} = 30$\,nm}}}}
\put(-4300,-150){\makebox(0,0)[lb]{\smash{{\SetFigFont{7}{8.4}{\rmdefault}{\mddefault}{\updefault} $\text{rms}^{\text{a}} = 307.8$\,nm}}}}
\put(-2700,-150){\makebox(0,0)[lb]{\smash{{\SetFigFont{7}{8.4}{\rmdefault}{\mddefault}{\updefault} $\text{rms}^{\text{b}} = 4069.7$\,nm}}}}
\put(-1100,-150){\makebox(0,0)[lb]{\smash{{\SetFigFont{7}{8.4}{\rmdefault}{\mddefault}{\updefault} $\text{rms}^{\text{c}} = 30.2$\,nm}}}}
\put(-4300,-300){\makebox(0,0)[lb]{\smash{{\SetFigFont{7}{8.4}{\rmdefault}{\mddefault}{\updefault} $\text{rms}^{\text{diff,a}} = 10^{3}\%$}}}}
\put(-2700,-300){\makebox(0,0)[lb]{\smash{{\SetFigFont{7}{8.4}{\rmdefault}{\mddefault}{\updefault} $\text{rms}^{\text{diff,b}} =  14000\%$}}}}
\put(-1100,-300){\makebox(0,0)[lb]{\smash{{\SetFigFont{7}{8.4}{\rmdefault}{\mddefault}{\updefault} $\text{rms}^{\text{diff,c}} = 0.6\%$}}}}
\end{picture}

\label{f:initialization_sans_turbulence}}
\bigskip


\subfigure[]{

\includegraphics[width=13.3cm]{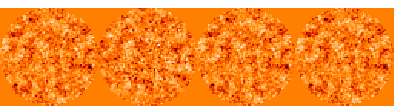}
\setlength{\unitlength}{4144sp}%
\begingroup\makeatletter\ifx\SetFigFont\undefined%
\gdef\SetFigFont#1#2#3#4#5{%
  \reset@font\fontsize{#1}{#2pt}%
  \fontfamily{#3}\fontseries{#4}\fontshape{#5}%
  \selectfont}%
\fi\endgroup%
\begin{picture}(0,0)
\put(-5950,1300){\makebox(0,0)[lb]{\smash{{\SetFigFont{7}{8.4}{\rmdefault}{\mddefault}{\updefault}``true''}}}}
\put(-4500,1300){\makebox(0,0)[lb]{\smash{{\SetFigFont{7}{8.4}{\rmdefault}{\mddefault}{\updefault}(a')}}}}
\put(-3000,1300){\makebox(0,0)[lb]{\smash{{\SetFigFont{7}{8.4}{\rmdefault}{\mddefault}{\updefault}(b')}}}}
\put(-1500,1300){\makebox(0,0)[lb]{\smash{{\SetFigFont{7}{8.4}{\rmdefault}{\mddefault}{\updefault}(c')}}}}
\put(-5800,-150){\makebox(0,0)[lb]{\smash{{\SetFigFont{7}{8.4}{\rmdefault}{\mddefault}{\updefault} $\text{rms}^{\text{``true''}} = 30$\,nm}}}}
\put(-4300,-150){\makebox(0,0)[lb]{\smash{{\SetFigFont{7}{8.4}{\rmdefault}{\mddefault}{\updefault} $\text{rms}^{\text{a'}} = 30.7$\,nm}}}}
\put(-2700,-150){\makebox(0,0)[lb]{\smash{{\SetFigFont{7}{8.4}{\rmdefault}{\mddefault}{\updefault} $\text{rms}^{\text{b'}} = 30.1$\,nm}}}}
\put(-1100,-150){\makebox(0,0)[lb]{\smash{{\SetFigFont{7}{8.4}{\rmdefault}{\mddefault}{\updefault} $\text{rms}^{\text{c'}} = 30.1$\,nm}}}}
\put(-4300,-300){\makebox(0,0)[lb]{\smash{{\SetFigFont{7}{8.4}{\rmdefault}{\mddefault}{\updefault} $\text{rms}^{\text{diff,a'}} = 180\%$}}}}
\put(-2700,-300){\makebox(0,0)[lb]{\smash{{\SetFigFont{7}{8.4}{\rmdefault}{\mddefault}{\updefault} $\text{rms}^{\text{diff,b'}} = 17\%$}}}}
\put(-1100,-300){\makebox(0,0)[lb]{\smash{{\SetFigFont{7}{8.4}{\rmdefault}{\mddefault}{\updefault} $\text{rms}^{\text{diff,c'}} = 17\%$}}}}
\end{picture}%

\label{f:initialization_turbulence}}\\

\bigskip

\caption{\textbf{Choice of an appropriate starting point}. Estimated upstream aberration maps with one spectral channel for three different starting points. [i] Without turbulent aberrations in the simulated images. [ii] With turbulent aberrations in the simulated images. From left to right, with a dynamic range adapted to the visualization: ``true'' simulated aberration map, (a) and (a') estimated aberrations with a random aberration map (rms value of the simulated aberrations) as starting point, (b) and (b') estimated aberrations with a zero aberration map as starting point, and (c) and (c') estimated aberrations with a random aberration map (rms value $10^{8}$ times lower than the ``true'' one) as starting point. The estimation is performed with a regularization on the star flux.}

\label{f:initialization}
\end{center}
\end{figure*}

We 
tested the phase retrieval capability of our algorithm with respect to the chosen starting point, assuming that there is no object to estimate. 
We give the rms value of the estimated upstream aberration map and the rms value of the difference between the simulated and the estimated maps estimated as follows:
 \begin{align}\label{eq:calcul_erreur_cartes_aberrations}
\text{rms}^{\text{diff}} = \frac{\Brack{\sum_{\rho} \Paren{\delta_u^\text{simulated}-\delta_u^\text{estimated}}^2}^{1/2}}{\Brack{\sum_{\rho} \Paren{\delta_u^\text{simulated}}^2}^{1/2}} \times 100.
 \end{align}
The inversion is performed with one spectral channel and without turbulence. 
Figure~\ref{f:initialization} compares some estimated upstream aberration maps (a, b, c) to the simulated one (``true''):
\begin{description}
\item[(a)] Using
a random aberration map with the same rms value as the true aberrations (30\,nm at 950\,nm) as a starting point does not help in finding
the global minimum. 
Indeed, the algorithm converges very quickly 
toward a local minimum and the estimated aberration map ($\text{rms}^{\text{a}}=307$\,nm at 950\,nm) is completely different from the simulated one ($\text{rms}^{\text{diff,a}}=10^{3}\%$). \color{black}
\item[(b)] Using a zero-aberration map as a starting point does not work either. 
This is probably 
because the approximate model is an even function. For this particular starting point, the gradient is null, which leads to some convergence difficulties. 
The rms of the difference between the two maps is about $1.4 \times 10^4\%$.
The estimated aberration pattern ($\text{rms}^{\text{b}}=4069$\,nm at 950\,nm) seems to show that the algorithm does not explore the high frequencies.
\item[(c)] The solution we propose is to use as a starting point for the minimization a non-null random aberration map with a 
low rms value compared to those of the ``true'' simulated aberration map. In practice, we 
chose an rms value about $10^{8}$ times 
lower than the ``true'' value. This leads to a correct estimation of the aberration map ($\text{rms}^{\text{c}}=30.2$\,nm at 950\,nm)  with an rms of the difference between the two maps of about 0.6\%.  
\end{description}
If we plot the same results for images simulated with turbulence,
the conclusions are not the same. The convergence to an aberration map that resembles the true one and has similar rms value 
is easier:
\begin{description}
\item[(a')] By using a random aberration map with the same rms value as the true aberrations (30\,nm at 950\,nm) as a starting point, the rms value of the estimated aberration map (30.7\,nm) is close to the true value. After a careful inspection, the estimated aberration map turned out to be similar to the opposite of the simulated one. This 
is discussed in more detail in~\ref{sss:4phases}.
\item[(b')] Using a zero-aberration map as a starting point leads to a good pattern and a good rms value of aberration map (30.1\,nm).
\item[(c')] Using a non-null random aberration map with a 
low rms value as a starting point also leads to a good pattern and a good rms value of the aberration map (30.1\,nm).
\end{description}
For the (b') and (c') cases, the difference between the two maps is bigger than before (17\%) but the estimated aberration map is a sufficiently good 
starting point for the alternating minimization.

The choice of an appropriate starting point seems not to be as essential with turbulent aberrations as it was without turbulent aberrations.
Indeed, the presence of turbulent aberrations raises the ambiguity that exists between the value of the star flux value and the rms value of the upstream aberration map.
Nevertheless, even if we assume that there are turbulent aberrations in the following, we 
chose to use a random aberration map with a 
low rms value as a starting point of the \textit{phase retrieval} 
because it allows us to avoid some local minima by linearizing the highly non-linear model used in the inversion.


\renewcommand{\thesubfigure}{\roman{subfigure}}
\captionsetup[subfigure]{labelformat=simple,labelsep=colon,listofformat=subsimple}

\renewcommand{\thesubfigure}{\alph{subfigure}}
\captionsetup[subfigure]{labelformat=simple,labelsep=colon,listofformat=parens}

 \subsubsection{Avoiding some local minima by testing quasi-equivalent starting points}\label{sss:4phases}
In the approximate model, four different aberration maps can give the same image (cf. Equation~(\ref{eq:phases_equivalentes}) in Appendix A). This means that, from a given starting point, the minimization algorithm can take four different but equivalent directions from the approximate model point of view. But from the point of view of the model used in the inversion, 
this is not the case because it depends on downstream aberrations, which break the symmetry. That is why we call them ``quasi-equivalent'' aberrations maps (cf. \ref{sss:approximate_model}).

The idea is then to explore the several regions offered by the four different 
quasi-equivalent aberrations maps to determine which of these solutions gives the smallest criterion. To do this, 
we 
performed an initialization step where the very small random phase is taken as a starting point. A first \textit{phase retrieval} stage 
was performed with this starting point, leading to a first estimated aberration map denoted by ${\delta_u}^{init,1}(\rho)$. Then, the three other quasi-equivalent aberration maps ${\delta_u}^{init,1}(-\rho)$, $-{\delta_u}^{init,1}(\rho)$ and $-{\delta_u}^{init,1}(-\rho)$ 
were taken as starting points for three other \textit{phase retrieval} stages. This 
led to three more estimated aberration maps denoted by ${\delta_u}^{init,2}$, ${\delta_u}^{init,3}$ and ${\delta_u}^{init,4}$.

\begin{figure*}
\begin{center}

\begin{picture}(400,95)%
\includegraphics[width=13.3cm]{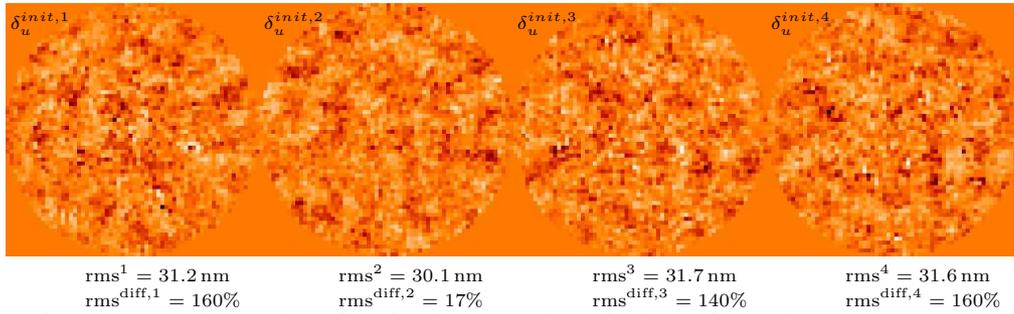}
\end{picture}%
\setlength{\unitlength}{4144sp}%
\begingroup\makeatletter\ifx\SetFigFont\undefined%
\gdef\SetFigFont#1#2#3#4#5{%
  \reset@font\fontsize{#1}{#2pt}%
  \fontfamily{#3}\fontseries{#4}\fontshape{#5}%
  \selectfont}%
\fi\endgroup%
\begin{picture}(0,0)
\put(-6300,1350){\makebox(0,0)[lb]{\smash{{\SetFigFont{7}{8.4}{\rmdefault}{\mddefault}{\updefault}$\delta_u^{init,1}$}}}}
\put(-4800,1350){\makebox(0,0)[lb]{\smash{{\SetFigFont{7}{8.4}{\rmdefault}{\mddefault}{\updefault}$\delta_u^{init,2}$}}}}
\put(-3300,1350){\makebox(0,0)[lb]{\smash{{\SetFigFont{7}{8.4}{\rmdefault}{\mddefault}{\updefault}$\delta_u^{init,3}$}}}}
\put(-1800,1350){\makebox(0,0)[lb]{\smash{{\SetFigFont{7}{8.4}{\rmdefault}{\mddefault}{\updefault}$\delta_u^{init,4}$}}}}
\put(-5900,-150){\makebox(0,0)[lb]{\smash{{\SetFigFont{7}{8.4}{\rmdefault}{\mddefault}{\updefault} $\text{rms}^{\text{1}} = 31.2$\,nm}}}}
\put(-4400,-150){\makebox(0,0)[lb]{\smash{{\SetFigFont{7}{8.4}{\rmdefault}{\mddefault}{\updefault} $\text{rms}^{\text{2}} = 30.1$\,nm}}}}
\put(-2900,-150){\makebox(0,0)[lb]{\smash{{\SetFigFont{7}{8.4}{\rmdefault}{\mddefault}{\updefault} $\text{rms}^{\text{3}} = 31.7$\,nm}}}}
\put(-1400,-150){\makebox(0,0)[lb]{\smash{{\SetFigFont{7}{8.4}{\rmdefault}{\mddefault}{\updefault} $\text{rms}^{\text{4}} = 31.6$\,nm}}}}
\put(-5900,-300){\makebox(0,0)[lb]{\smash{{\SetFigFont{7}{8.4}{\rmdefault}{\mddefault}{\updefault} $\text{rms}^{\text{diff,1}} = 160\%$}}}}
\put(-4400,-300){\makebox(0,0)[lb]{\smash{{\SetFigFont{7}{8.4}{\rmdefault}{\mddefault}{\updefault} $\text{rms}^{\text{diff,2}} = 17\%$}}}}
\put(-2900,-300){\makebox(0,0)[lb]{\smash{{\SetFigFont{7}{8.4}{\rmdefault}{\mddefault}{\updefault} $\text{rms}^{\text{diff,3}} = 140\%$}}}}
\put(-1400,-300){\makebox(0,0)[lb]{\smash{{\SetFigFont{7}{8.4}{\rmdefault}{\mddefault}{\updefault} $\text{rms}^{\text{diff,4}} = 160\%$}}}}
\end{picture}\\

\bigskip

\caption{\textbf{Estimated upstream aberrations for the four quasi-equivalent aberration maps as starting points.} From left to right, with the same dynamic range: $\delta_u^{init,1}$,  $\delta_u^{init,2}$, $\delta_u^{init,3}$, $\delta_u^{init,4}$. The image simulation is performed with one spectral channel in the presence of turbulent aberrations.}
\normalsize
\label{f:4phases_turbulence}
\end{center}
\end{figure*}

Figure~(\ref{f:4phases_turbulence}) shows the four estimated aberration maps
at the end of the initialization step. These estimated aberration maps are compared to the simulated one (Fig. (\ref{f:initialization}.d)). 
The final aberration map chosen as a starting point for the alternating algorithm is the one that gives the minimum value for criterion J of Equation \ref{critere}:
\begin{align}
 \Paren{\delta_u}_{init} = \text{arg min} \Brace{     J\Brack{    \delta_u^{init,1}    }, J\Brack{   \delta_u^{init,2}     }, J\Brack{    \delta_u^{init,3}     },J\Brack{      \delta_u^{init,4}       }         }. 
\end{align}
In practice, quite often and in particular in this simulation, it turns out that the chosen set of aberrations is also the one 
whose rms value ($\text{rms}^{\text{b'}}=30.1$\,nm at 950\,nm) is closest 
to the `true'' phase (Fig. (\ref{f:initialization_sans_turbulence}. ``true'') and Fig. (\ref{f:initialization_turbulence}. ``true'')).

\subsubsection{Avoiding some local minima in the multispectral inversions by taking the previously estimated aberration map as starting point}\label{sss:}
We 
used a local descent algorithm to minimize the criterion. 
For this, the gradients are computed for each explored direction and the computation is all the longer as there are spectral channels. 
That is why it is useful to perform several inversions by gradually increasing the number of spectral channels. 
This also helps us avoid some local minima.
We begin by an inversion with one spectral channel. Then, we add some more spectral channels for new inversions and each time, we take the previous estimated aberration map as a starting point. 


Because an inversion with only one spectral channel
sometimes leads to a local minimum, 
it is useful to also test the four quasi-equivalent starting points 
with two spectral channels (cf. Section~\ref{sss:4phases}). 

\subsection{VMLM optimizer}\label{VMLM}
To minimize the criterion, we 
chose the variable metric with limited memory and bounds (VMLM-B) method~\citep{Thiebaut-p-02}. 
Updated from the BFGS \emph{variable metric} method~\citep{Numerical-Recipes-3}, it is usable for a problem of large dimensionality. 
Moreover, it offers the possibility to constrain these parameters. 
This makes this method 
a good tool for many inversion problems in high angular resolution~\citep{meimon-a-09,Gratadour-a-05}. 
It is available from \url{http://www-obs.univ-lyon1.fr/labo/perso/eric.thiebaut/optimpack.html}.

\subsection{Summary of the developed algorithm}
\begin{figure*}
\begin{center}
\includegraphics[angle= -90, width=18cm]{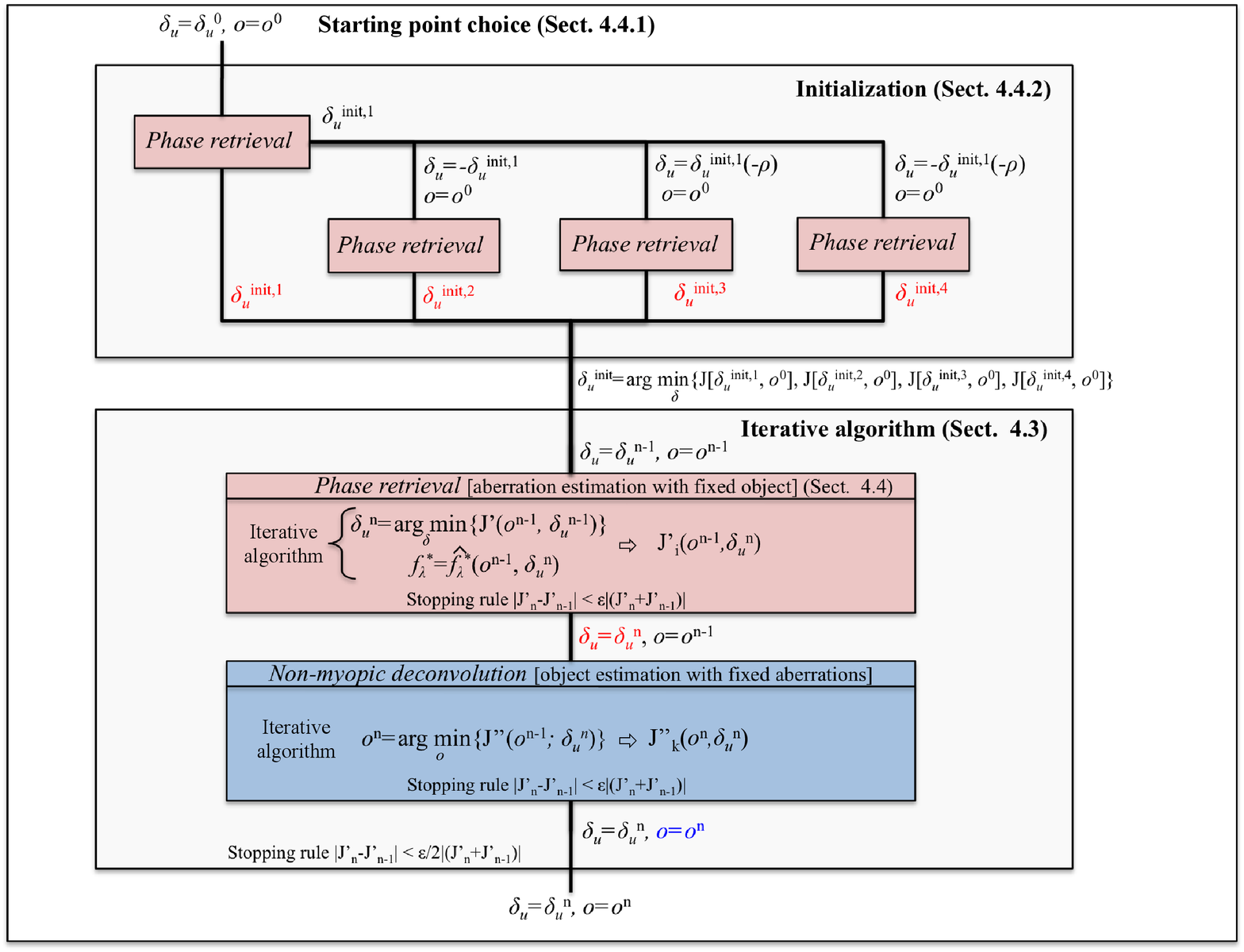}
\caption{\textbf{Block diagram of the algorithm used for the joint estimation of the object map and the upstream aberrations.}}
\normalsize
\label{f:bloc_diagram}
\end{center}
\end{figure*}

Figure~(\ref{f:bloc_diagram}) summarizes the different steps of the developed algorithm.
The choice of a very small random phase as a starting point is essential because it avoids falling into some local minima (Section \ref{sss:starting_point}). An initialization phase is performed. 
It consists in running the algorithm for the four quasi-equivalent solutions (Section \ref{sss:4phases}). The solution 
that leads to the 
lowest criterion value is selected. Then, the minimization core is performed, alternating between the aberration estimation, assuming that the object is known (multispectral \emph{phase retrieval}), and the object estimation, assuming that the aberrations are known (\emph{non-myopic multispectral deconvolution}). 
Regularization terms and constraints prevent the algorithm from falling into other local minima (Section \ref{regul_flux}).
Iterations are performed (Section \ref{iterative_algorithm}) until the stopping rule of the algorithm is verified. The chosen optimizer is the VMLM-B~\citep{Thiebaut-p-02} (Section \ref{VMLM}).

\section{Validation of the inversion method by simulations}\label{s:validations}

In this section, we validate the exoplanet detection capabilities of our inversion method. After giving the numerical simulation conditions, we investigate the estimation quality of the aberrations and the object as a function of the number of images at different wavelengths used. 
We also study the algorithm robustness with respect to the simulated images and with respect to the starting point we use.
Finally, we study the effect of the bandwidth on the quality of the object estimation.

\subsection{Simulation hypothesis}\label{ss:hyp}
\begin{figure}
\begin{center}
\subfigure[Object map]{
\includegraphics[width=3.75cm]{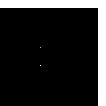}
\label{f:objet_simule}}
\subfigure[Image of the object map in the focal plane]{
\includegraphics[trim= 0mm 0mm 20.25mm 0mm, clip, width=3.75cm]{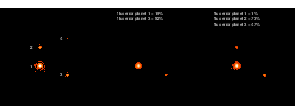}
\label{f:image_objet_simule}}\\
\subfigure[Aberration map]{
\includegraphics[width=3.75cm]{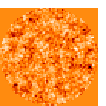}
\label{f:aberrations_simulees}}
\subfigure[Image of the speckle field in the focal plane]{
\includegraphics[width=3.75cm]{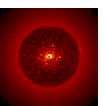}
\label{f:image_aberrations_simulees}}
\caption{\textbf{Simulated images at $\lambda=950$\;nm.} (a) Simulated object map and (b) associated image in the focal plane. The following planets are simulated: one with a star-over-planet contrast of $10^5$ at a separation of 0.2 arcsec (planet 1), two with star-over-planet contrast of $10^6$ at separations of 0.2 (planet 2) and 0.4 (planet 3), respectively, and one with a star-over-planet contrast of $10^7$ at a separation of 0.4 arcsec (planet 4). The image in the focal plane is obtained by convolving the object map $o_{\lambda}$ by the non-coronagraphic psf $h_{\lambda}^{nc}$. (c) Simulated aberrations and (d) associated image of the speckle field in the image focal plane. The image is given by the 
coronagraphic PSF $h_{\lambda}^{c}$.}
\label{f:image_simulees}
\end{center}
\end{figure}

\begin{figure*} 
\begin{center}
\subfigure[Power spectral density of the upstream aberrations.]{
\includegraphics[width=6.3cm]{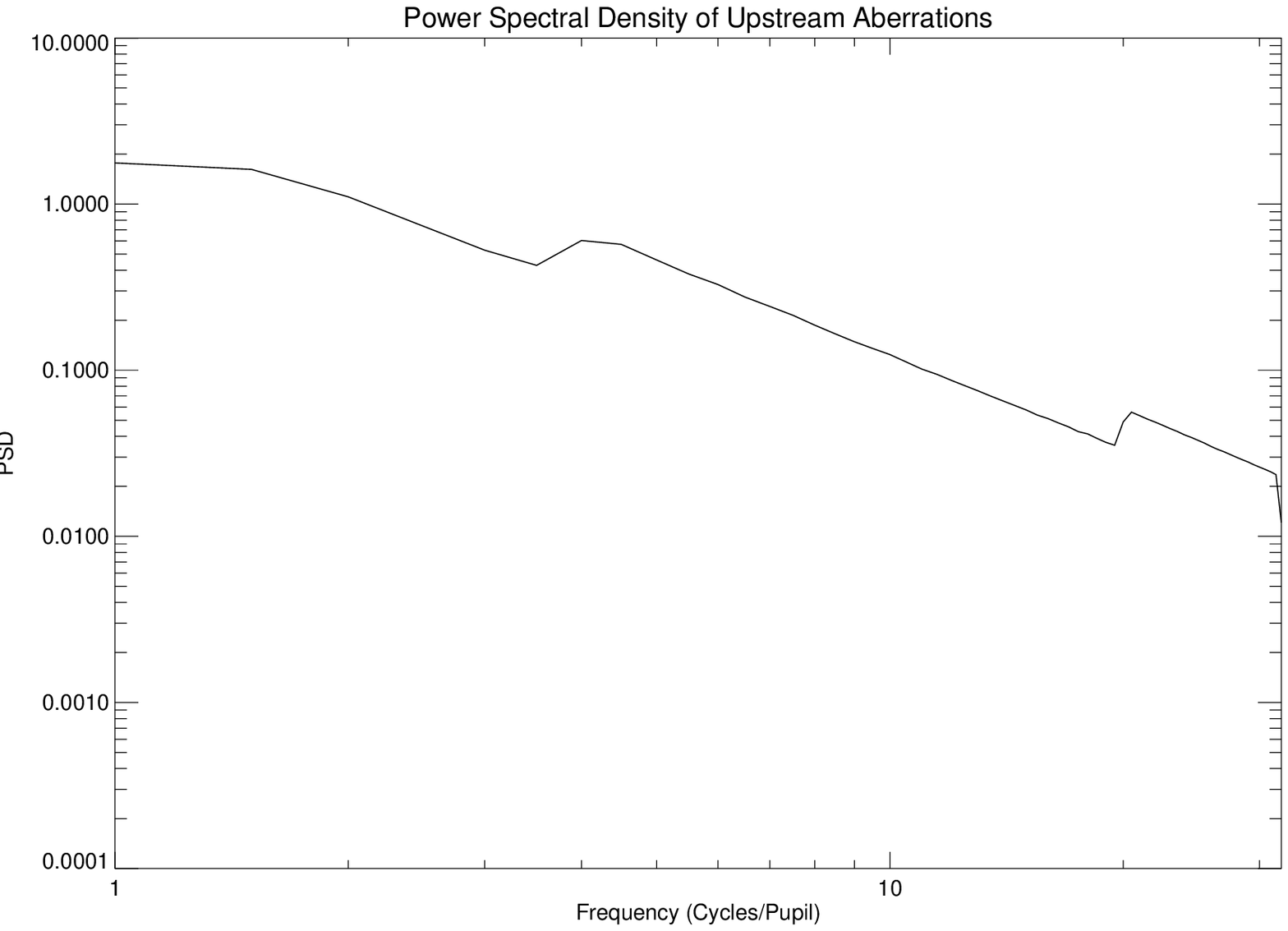}
\label{sf:dsp_phi_up}}
\subfigure[Power spectral density of the downstream aberrations.]{
\includegraphics[width=6.3cm]{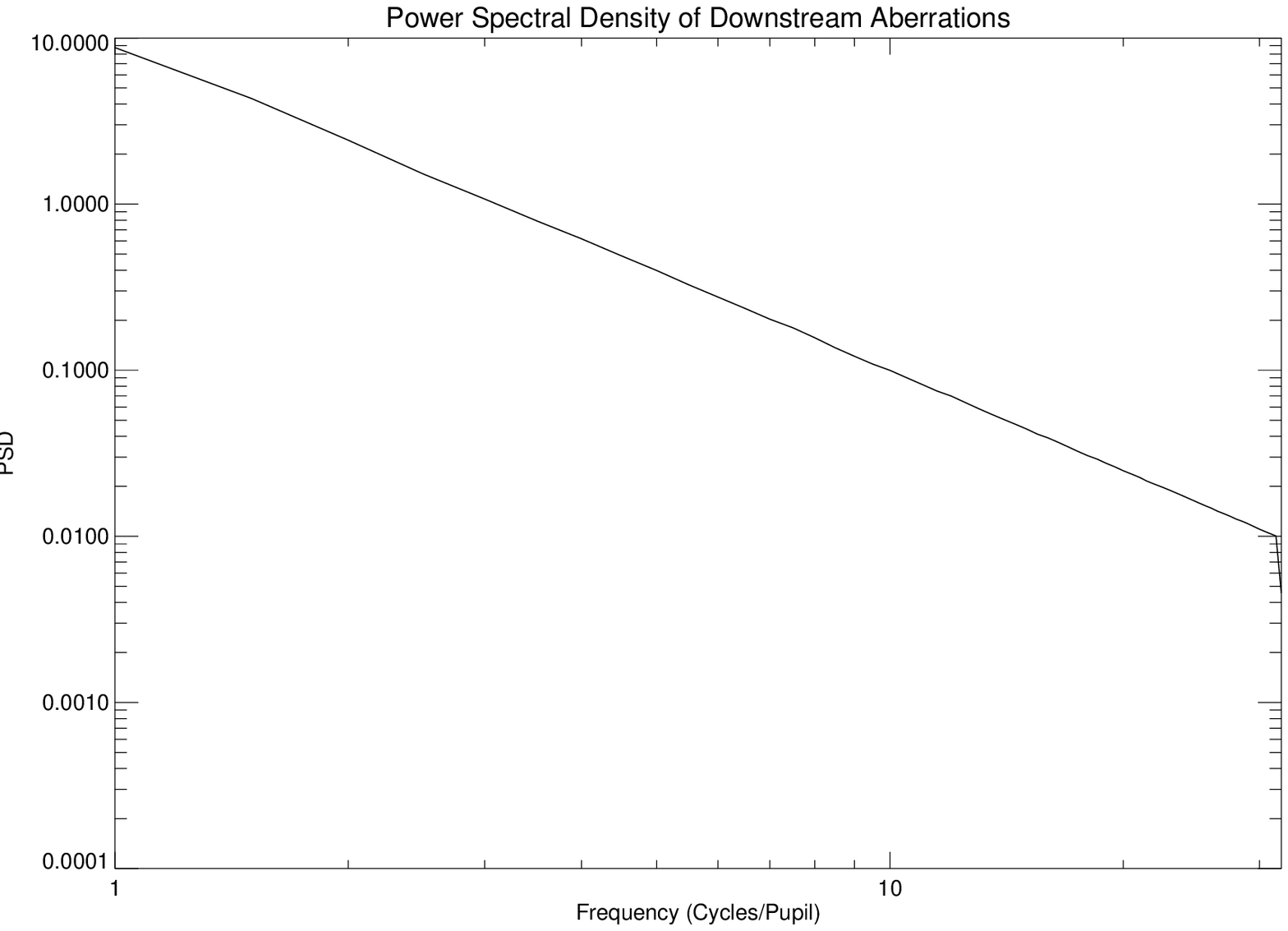}
\label{sf:dsp_phi_do}}
\caption{\textbf{Power spectral densities of the simulated aberrations.} The upstream and downstream aberrations are randomly generated according to a $f^{-2}$ spectrum and with an rms value of 30\,nm. The downstream aberrations are not corrected by the adaptive optics. The upstream aberrations are corrected by the adaptive optics up to 20 cycles/pupil. The residuals are due to the non-common path aberrations. Below 4 cycles/pupil, these non-common path aberrations are corrected but there are some residuals from rotative optics.}
\label{f:dsp}
\end{center}
\end{figure*}

From a data cube of six images simulated with the image formation model of equation~(\ref{imaging_model}) and the \citet{Sauvage-a-10} analytical expression for coronagraphic imaging (cf. Equation~(\ref{model_jeff}) in Appendix B), we jointly 
estimated the speckle field and the object map. 
We 
chose pixel indicator functions as the basis for the phase rather than, e.g., a truncated basis of Zernike polynomials, 
to model and reconstruct phases with a high spatial frequency content. 
The hypotheses are typical of a SPHERE-like instrument: upstream $\delta_u$ and downstream $\delta_d$ aberrations simulated with standard deviation of 30\,nm (cf. their power spectral densities in Figure~(\ref{f:dsp})), star-planet angular separations of 0.2 and 0.4\,arcsec, contrasts, i.e. ratio of star flux over planet flux of $10^5$, $10^6$ and $10^7$, a [950\,nm;1647\,nm] spectral bandwidth and an integrated flux of $4 \times 10^{11}$ 
on the data cube in presence of photon noise and a transmission (throughput and quantum efficiency) of 10\%, corresponding to the observation of a 6-magnitude star for 25 minutes with the VLT.
We 
used $128 \times 128$ pixels to simulate our images, Shannon-sampled at 950\,nm. 
This results in a number of unknowns to estimate for the aberration map 
of about $3 \times 10^3$.
If we add the unknowns to estimate for the object map, which is  $16 \times 10^3$, the total number of unknowns is about $2 \times 10^4$.
Figure (\ref{f:image_simulees}) shows the simulated objet map~(\ref{f:objet_simule}) and the associated image in the focal plane~(\ref{f:image_objet_simule}). 
For an easier visualization, we represent the images in the focal plane and not the object map in the following. 
Figure \ref{f:image_simulees} shows the simulated aberration map~(\ref{f:aberrations_simulees})  and the associated image of the speckle field in the focal plane~(\ref{f:image_aberrations_simulees}).

\subsection{Algorithm robustness and performance studies}
We jointly 
estimated the upstream quasi-static aberration map and the object map with multispectral data.  
To study the robustness of the method we have developed, we 
ran several simulations in a 
Monte Carlo-like manner.
Both different simulated images (\ref{sss:different_phases}) and different starting point (\ref{sss:different_starting_points}) 
were used for the inversion.
The results of the inversions with two and six spectral channels 
were compared 
to study the effect of the redundancy of information offered by the multispectral imaging.


\subsubsection{Different simulated phases}\label{sss:different_phases}

\begin{figure*}
\begin{center}
\subfigure[two-spectral channel inversion]{
\includegraphics[trim= 0mm 19mm 0mm 0mm, clip, width=13cm]{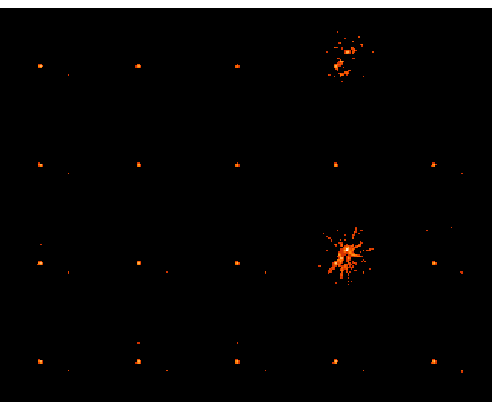}
\label{sf:robustesse_phases_2lambda}}
\subfigure[six-spectral channel inversion]{
\includegraphics[trim= 0mm 0mm 0mm 19mm, clip, width=13cm]{2012722_robustesse_phases}
\label{sf:robustesse_phases_6lambda}}\\
\caption{\textbf{Robustness study on the simulated phases.} With the same dynamic range, at 950nm: estimated planet images $\Brack{o_\lambda\star h_\lambda^{nc}}\Paren{x,y}$ from different images, simulated with ten different randomly generated upstream aberration maps.}
\label{f:robustness_study}
\end{center}
\end{figure*}

We 
applied our method to ten images simulated with different random upstream aberration maps to assess the algorithm robustness. 
Figure~\ref{f:robustness_study} shows the estimated images of the object from these images for a 
two-spectral channel inversion (a) and a 
six-spectral channel inversion (b). 
These results bring several conclusions, both on the robustness of the method and the multispectral redundancy.

\paragraph{Robustness of the method.}
In 
one out of 
ten cases, some residuals from the speckle field remain on the object. The level of this residual 
prevents one from detecting any planet. In the other 
nine cases, the level of the residuals is negligible, which allows one to detect at least one planet. These observations, independent 
of the number of spectral channels used for the inversion, shows that the method is relatively robust, given the minimization difficulties we met.

\begin{figure*}
\begin{center}
\includegraphics[width=13cm]{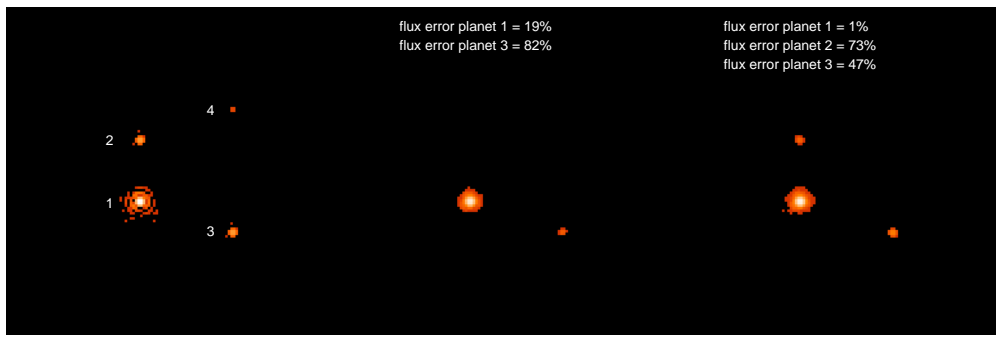}
\caption{\textbf{Multispectral redundancy.} With the same dynamic range, at 950nm, from left to right: simulated planet image $\Brack{o_\lambda\star h_\lambda^{nc}}\Paren{x,y}$ and estimated ones with a two- and six-spectral channel inversion. The errors on the planet flux estimations are computed when the planet is detected. }
\label{f:estimation_flux}
\end{center}
\end{figure*}

\paragraph{Multispectral redundancy.}
We can see some significant differences between the 
two-spectral channel inversion and the 
six-spectral channel inversion:
\begin{itemize}
\item The planet with a contrast of $10^5$ is detected in 
eight out of 
ten cases with two spectral channels and in 
nine out of 
ten cases with six spectral channels.
\item Only one planet with a contrast of $10^6$ is detected with two spectral channels in 
three out of 
ten cases. The same planet is detected in 
nine out of 
ten cases with six spectral channels.
\item The other planet with a contrast of $10^6$ is detected in 
three out of 
ten cases with six spectral channels.
\item Furthermore, the estimation of the flux of the planets is 
more accurate with six spectral channels. The inversion shown in Figure~\ref{f:estimation_flux} is representative of a large body of simulated tests and is performed with two and six spectral channels. The simulated and estimated image of the object maps for these two different inversions are represented as the errors on the estimated planet flux when the planet is effectively detected.
\end{itemize}
These results show that multispectral redundancy helps us detect the planets.
However, we note that even if the convergence problems were solved 
with our minimization strategy in most cases, there are still challenges to 
be overcome to arrive at a perfectly robust algorithm.

\subsubsection{Different starting points}\label{sss:different_starting_points}

\begin{figure*}
\begin{center}
\includegraphics[width=13cm]{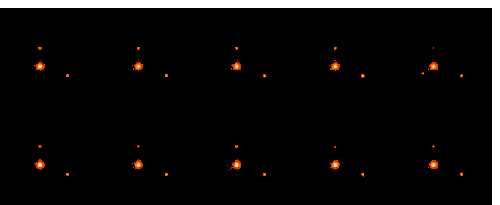}
\caption{\textbf{Robustness study on the aberrations maps used as starting points.} With the same dynamic range, at 950nm: estimated planet images $\Brack{o_\lambda\star h_\lambda^{nc}}\Paren{x,y}$ from one image with different random aberration maps used as starting points.}
\label{f:robustness_study2}
\end{center}
\end{figure*}

For each of the ten previous simulated phases, we 
took ten different random aberration maps with the same level of rms value as starting points to demonstrate than any random aberration map, provided it is small, allows us to find a good solution.
Figure~\ref{f:robustness_study2} shows the estimated images of the object for the ten different random aberration maps, used as starting points, for one phase.

The starting point does not have a 
strong impact on the planet detection. In all cases, the three 
brightest planets are detected. In only 
one out of 
ten cases, a false planet is detected. 
Despite all the attention given to the initialization of the algorithm, the choice of starting point can have, to a limited extent, an effect on detection.


%


\subsection{Bandwidth effect}
\begin{figure*}
\begin{center}
\includegraphics[width=16cm]{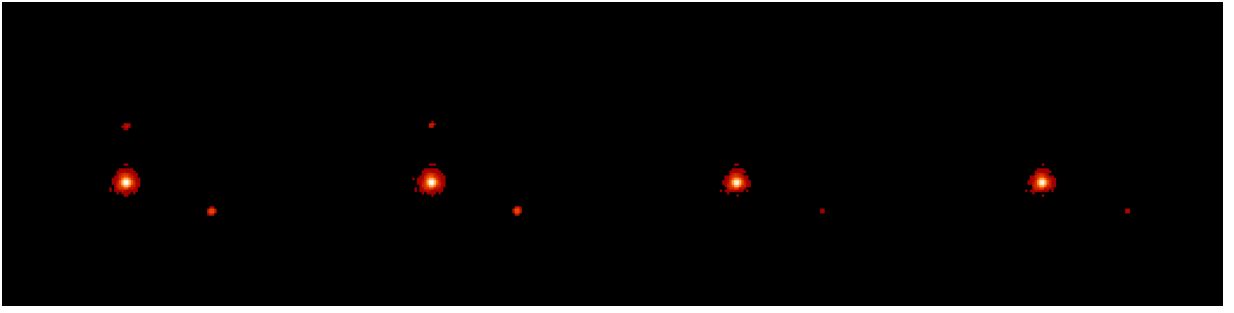}
\caption{\textbf{Bandwidth effect.} With the same dynamic range, at 950nm, from left to right: estimated planet images $\Brack{o_\lambda\star h_\lambda^{nc}}\Paren{x,y}$ with a 700, 400, 200, and 100\,nm-bandwidth inversion.}
\label{f:largeur_bande}
\end{center}
\end{figure*}

We study here the bandwidth effect on the planet image estimation quality. We 
selected the following spectral bandwidths: [950\,nm;1050\,nm], [950\,nm;1150\,nm], [950\,nm;1350\,nm], and [950\,nm;1650\,nm]. The two last bandwidths correspond to operational modes of the SPHERE-IFS instrument, whereas the other bandwidths are produced for the purpose of our study's completeness. 
Figure~(\ref{f:largeur_bande}) compares the estimated planet image maps for these different inversions. 
If the planet with a contrast of $10^5$ is detected each time, only the two 
broader bandwidths (700 and 400\,nm) allow one to detect the two planets with a contrast of $10^6$. 
With a 200 and a 100\,nm bandwidth, the planet with a contrast of $10^6$ which is at a separation of 0.4\,arcsec is detected, whereas the one 
that is at a separation of 0.2\,arcsec is not. 

The detection performance increases with the bandwidth because the incorporated spectral information helps the algorithm to distinguish between speckles and planets.
Indeed, from one wavelength to another, the amplitude of a speckle movement is proportional to the wavelength difference and to the radial position of the speckle.
For a given spectral bandwidth, it is 
therefore easier to detect a planet that is far away from the star than a planet than is close to the star.
\color{black}


%
%
%
%

\color{black}

\section{Conclusion}

We have proposed an original method 
that jointly estimates 
the object (multispectral deconvolution) and the aberrations (multispectral phase retrieval) for the new generation of planet finders. For the first time, a fine parametric model of coronagraphic imaging, describing the instrument response, is used for the inversion of simulated multispectral images in a solid statistical framework. 
Even though the model 
remains a simplification of reality, in particular when assuming achromatic wavefront errors, it goes much further than only assuming that the spatial speckle pattern essentially scales with wavelength. 
We have shown that the second-order approximation of the imaging model has the same behavior as the often-used model for the problem at hand: the speckle pattern is centrosymmetric, it scales with the wavelength, and it is not dependent on the downstream aberrations. 
The departure from this case 
quickly becomes very significant as the phase grows. 
It is clear that the ability of a finite phase to produce non-symmetric speckles induces a strong ambiguity between the estimates of the aberrations and the object, even if the latter is a point-like companion. 

To set up our method, we 
developed an iterative algorithm. With only one spectral channel, this joint estimation is an underdetermined problem, as 
we emphasized before. This underdetermination results mathematically in a degeneracy of the global minimum. 
A multispectral inversion raises this underdetermination but it is still possible to fall into local minima. Because of the high non-linearity of the coronagraphic imaging analytical model and the number of unknowns to estimate (about $10^3$ in our case), the phase retrieval, even if it is multispectral, remains a difficult 
problem. 
We 
set-up a minimization approach that remains quite fast (without systematically exploring 
the whole parameter space in a search for global minimum) but that is still relatively robust: extensive tests 
showed the success 
in converging to 
a good solution in 90\% of the cases in a systematic manner, and in the other cases, the failure appears obviously in the results with no risk of confusion and can motivate tests with alternative criterion minimization approaches.  
We 
obtained 
these convergence capabilities of the algorithm by bringing original solutions to the minimization difficulties of the phase retrieval, inspired 
by studying the imaging model. One element of the solution is to use a small random aberration map as starting point. Another 
element is to explore the several directions offered by the 
quasi-equivalent aberration maps.


A wide variety of prior information, either about the system (aberrations, flux, noise) or about the object of interest, can be used to constrain the problem. The choice of a Bayesian approach allows this flexibility. In particular, a regularization 
that minimizes the energy on the object map 
helped us in separating 
the aberrations from the object and 
in decreasing the speckle noise in the reconstructed object map.


The restoration of images simulated with a perfect coronagraph is very encouraging for the extraction of planetary signals at levels 
that begin to be astrophysically interesting. We 
demonstrated the efficiency of the method even with only two spectral channels, by achieving a contrast of $10^5$ at 0.2\,arcsec. Multispectral redundancy improves the detection, 
which allowed us to achieve a contrast of $10^6$ at 0.2\,arcsec with six spectral channels. 


We 
therefore believe that this approach 
will be quite powerful when we are faced with experimental data.
This deserves to be studied, as well as how the performance will evolve in the cases of images simulated with a non-perfect coronagraph, real images from the SPHERE instrument in the lab, or real images from an instrument on-sky.
Eventually, this method could be used to improve the performance of the existing multispectral imaging instruments, providing better astrophysical exploitations. 
Now that we demonstrated that we 
can manage the difficulties linked to the criterion minimization,
we can now focus on its applicability, and 
on adding more 
prior information, starting with the full set of information that can be obtained from the instrument calibrations. %


The lessons 
learned by applying the method could also facilitate the approach for the design of future instruments such as EPICS for the European Extremely Large Telescope~\citep{Kasper-p-08}, and the definition of their calibration procedures.

\section*{Appendix A: Indetermination on the estimated aberrations from an image simulated with our approximate model}\label{a:Appendix_A}

We show here that four sets of upstream aberrations give the same image 
for our approximate model. 
We re-write the expression 
below as a function of $\phi_u = (2 \pi / \lambda) \times \delta_u$ and without the variables for 
better readability:
\begin{align}\label{eq:approximate_model}
\Brack{h_{\lambda}^c}^{app}
&= \abs{\wt{\mathcal{P}_d } \star \wt{\phi_u }}^2 \nonumber\\
&+   \abs{\wt{\mathcal{P}_d}}^2 \star S_{\phi_r} \Paren{\alpha} - \moy{\abs{P\Brack{\phi_r}}^2}_t \cdot \abs{\wt{\mathcal{P}_d}}^2  \nonumber\\
&+  o\Paren{\phi^2}.
\end{align}

We consider here the static term $ \abs{\wt{\mathcal{P}_d } \star \wt{\phi_u }}^2$ and re-write it 
in the form of a correlation. For this, we consider two functions  $f = \mathcal{P}_d$ and  $g = \delta_u$ of the two variables $\rho_x$ and $\rho_y$ and 
denote $\check{f}(\rho) = f(-\rho)=   f(-\rho_x, -\rho_y)$. 

By using the definition of the correlation $\Gamma_{fg}(\rho)$ and the convolution $\mathcal{C}_{fg}(\rho)$ of the two functions $f(\rho)$ and $g(\rho)$, 
\begin{align}
\Gamma_{fg}(\rho) &= f(\rho) \otimes g(\rho) = \int f^*(\rho') g(\rho'+\rho) \ddroit \rho, \nonumber\\
\mathcal{C}_{fg}(\rho) &= f(\rho) \star g(\rho) = \int f(\rho') g(\rho-\rho') \ddroit \rho, \nonumber
\end{align}
and the 
properties 
\begin{align}
\Paren{\wt{f}}^* = \wt{\check{f}^*} \text{   and   } f \star g = f \otimes \check{g}^*, \nonumber
\end{align}
we 
obtain
\begin{align}
\abs{\wt{f} \star \wt{g}}^2 = \wt{fg} \cdot (\wt{fg})^* = \wt{fg} \cdot \wt{\check{(fg)^*}} =  \wt{fg \star \check{(fg)^*}} = \wt{fg \otimes fg} =  \wt{\Gamma_{fg}}. \nonumber
\end{align}
This yields
\begin{align}
\abs{\wt{\mathcal{P}_d} \star \wt{\phi}_u}^2 = \wt{\Gamma_{(\mathcal{P}_d \cdot \phi_u)}}.
\end{align}

The 
properties of the autocorrelation 
\begin{align}
\Gamma_{f(\rho)} &= \Gamma_{f^*(-\rho)} \nonumber
 \intertext{and}
\Gamma_{-f(\rho)} &= \Gamma_{f(\rho)}, \nonumber
\end{align}
lead 
to 
\begin{align}
& \left|
  \begin{array}{l l l}
     \Gamma_{(\mathcal{P}_d \cdot \delta_u)(\rho)}  &=  \Gamma_{(\mathcal{P}_d \cdot \delta_u)(-\rho)}\\
    \Gamma_{(\mathcal{P}_d \cdot (-\delta_u))(\rho)}  &= \Gamma_{(\mathcal{P}_d \cdot (-\delta_u))(-\rho)} \\
  \end{array} \right. \nonumber
 \intertext{and}
& \Gamma_{(\mathcal{P}_d \cdot (-\delta_u))(\rho)} = \Gamma_{(\mathcal{P}_d \cdot \delta_u)(\rho)} .\nonumber
\end{align}
This yields
\begin{align}\label{eq:phases_equivalentes}
\Gamma_{(\mathcal{P}_d \cdot \delta_u)(\rho)} = \Gamma_{(\mathcal{P}_d \cdot \delta_u)(-\rho)} = \Gamma_{(\mathcal{P}_d \cdot (-\delta_u))(\rho)}= \Gamma_{(\mathcal{P}_d \cdot (-\delta_u))(-\rho)}, 
\end{align}
which means that the upstream aberration sets $\delta_u(\rho)$, $\delta_u(-\rho)$, $-\delta_u(\rho)$ and $-\delta_u(-\rho)$ are equivalent with respect to the approximate model, because they give the same image.
This is true even in the presence of the turbulent term $\abs{\wt{\mathcal{P}_d}}^2 \star S_{\phi_r} \Paren{\alpha} - \moy{\abs{P\Brack{\phi_r}}^2}_t \cdot \abs{\wt{\mathcal{P}_d}}^2$.


\section*{Appendix B: Indetermination on the estimated aberrations from an image simulated with the Sauvage et al. model}\label{a:Appendix_B}
In classical imaging, i.e. ``non-coronagraphic imaging'', the sign of the even part of the phase cannot be deduced from only one image in the focal plane \citep{Blanc-t-02}. In other words, if we denote $\phi_e$ and $\phi_o$, the even and the odd parts of the phase, the two phases 
$\phi = \phi_e + \phi_o$ and $\phi' = -\phi_e + \phi_o$ give the same image.
In this appendix, we show that 
this is also the case for the Sauvage et al. expression \citep{Sauvage-a-10} in coronagraphic imaging, if we assume that the sign of the even part changes for all 
phase errors.

The expression of Sauvage et al. is 
\begin{equation}\label{model_jeff}
  h_{\lambda}^c = \moy{A_nA_n^*} + \moy{\left| (\eta_0) \right|^2} A_dA_d^* - 2\Re\Brace{ \moy{\eta_0A_n^*}A_d },
\end{equation}
with   $A_n \Paren{\alpha} = \text{TF}^{-1}\Brack{ \mathcal{P}_d \Paren{\rho} e^{j\phi_{tot} \Paren{\rho}} }$,
  $A_d \Paren{\alpha} = \text{TF}^{-1}\Brack{ \mathcal{P}_d \Paren{\rho} e^{j\phi_d \Paren{\rho} }}$,
  $\phi_i \Paren{\rho} = 2\pi\frac{\delta_i\Paren{\rho}}{\lambda}$ and
  $\phi_{tot} \Paren{\rho} = \phi_r\Paren{\rho} + \phi_u\Paren{\rho} + \phi_d\Paren{\rho}$.
$\text{TF}\Brack{.}$ denotes the Fourier Transform. $\moy{\left| (\eta_0) \right|^2} $ represents the mean Strehl ratio during observation, such as 
\begin{align}
\eta_0(t) &= \moy{ \Psi_0(\rho)|\mathcal{P}_u(\rho)} \nonumber\\
&= \frac{1}{S} \iint_{\rho} \Psi_0^*(\rho)\mathcal{P}_u(\rho) {\ddroit}^2\rho \nonumber\\
&= \frac{1}{S^2}\iint_{\rho} \mathcal{P}_u^2(\rho) e^{-j\phi(\rho,t)} {\ddroit}^2\rho, \nonumber
\end{align}
with $\phi(\rho,t) = \phi_r(\rho,t)+\phi_u(\rho)$.

The first term of Sauvage et al.'s expression $\moy{A_nA_n^*}$ is the classical case of non-coronagraphic PSF, which is well-known \citep{Blanc-t-02}. The term $A_nA_n^*$ stays identical whatever the sign of the even part of the phase. 

The second term of Sauvage et al.'s expression is the product of two 
factors: $\moy{\left| (\eta_0) \right|^2}$ and $A_dA_d^*$. The latter stays identical whatever the sign of the even part of the phase. We take the following phase 
$\phi' = -\phi_e + \phi_o$ and 
calculate the corresponding $\eta_0'(t)$, assuming that $\rho'' = -\rho$:
\begin{align}
\eta_0'(t)&= \frac{1}{S^2}\iint_{\rho} \mathcal{P}_u^2(\rho) e^{-j\phi'\Paren{\rho,t}} {\ddroit}^2\rho  \nonumber\\
&= \frac{1}{S^2}\iint_{\rho} \mathcal{P}_u^2(\rho) e^{-j\Brack{-\phi_e\Paren{\rho,t}+\phi_o\Paren{\rho,t}}} {\ddroit}^2\rho  \nonumber\\
&= \frac{1}{S^2}\iint_{\rho} \mathcal{P}_u^2(\rho) e^{j\Brack{\phi_e\Paren{-\rho,t}+\phi_o\Paren{-\rho,t}}} {\ddroit}^2\rho  \nonumber\\
&= \frac{1}{S^2}\iint_{\rho''} \mathcal{P}_u^2(\rho'') e^{j\Brack{\phi_e\Paren{\rho'',t}+\phi_o\Paren{\rho'',t}}} {\ddroit}^2\rho''  \nonumber\\
&= \frac{1}{S^2}\iint_{\rho''} \mathcal{P}_u^2(\rho'') e^{j\Brack{\phi\Paren{\rho'',t}}} {\ddroit}^2\rho'' \nonumber\\
&= \Brack{\eta_0(t)}^*. \nonumber
\end{align}
$\moy{\left| (\eta_0) \right|^2} =  \moy{\left| (\eta_0 \cdot \eta_0(t)^*) \right|^2}$ is then independent of the sign of the even part of the phase. Thus, the product $\moy{\left| (\eta_0) \right|^2} A_dA_d^*$ is also independent of the sign of the even part of the phase.

We study now the third term $2\Re\Brace{ \moy{\eta_0A_n^*}A_d }$. Assuming that $\phi_d = (\phi_d)_e + (\phi_d)_o$  and $\phi'_d = -(\phi_d)_e + (\phi_d)_o$:
\begin{align}
A'_d \Paren{\alpha} &= \text{TF}^{-1}\Brack{ \mathcal{P}_d \Paren{\rho} e^{j\phi'_d \Paren{\rho} }} \nonumber\\
&= \iint_{\rho'} \Brack{ \mathcal{P}_d \Paren{\rho} e^{j\Brack{-(\phi_d)_e\Paren{\rho}  + (\phi_d)_o\Paren{\rho} } }} e^{-2i\pi\Paren{\rho\alpha}} {\ddroit}^2\rho \nonumber\\
&= \iint_{\rho'} \Brack{ \mathcal{P}_d \Paren{\rho''} e^{-j\Brack{(\phi_d)_e\Paren{\rho''}  + (\phi_d)_o\Paren{\rho''} } }} e^{2i\pi\Paren{\rho''\alpha}} {\ddroit}^2\rho'' \nonumber\\
&= \Brack{A_d}^*. \nonumber
\end{align}
In the same way as the previous demonstration, we can show that $A_n^* = A_n$.
Under the effect of the 
transformation 
$\phi \rightarrow \phi'$, the different terms become 
\[ \left\{ \begin{array} {r@{\quad}l}
\eta_0 \rightarrow \eta_0^* \\ A_n^* \rightarrow A_n \\ A_d \rightarrow A_d^*.
\end{array}\right.\] \\ 
In other words, $\moy{\eta_0A_n^*}A_d \rightarrow \Brack{\moy{\eta_0A_n^*}A_d}^*$
When we take the conjugate of a complex number, only the sign of the imaginary part changes.
Because we take the real part of this expression, changing the sign of the even part of the phase does not change the term.

To conclude, like in classical imaging, changing the sign of the even part of the phase does not change the image in the focal plane. This means that two sets of aberrations give the same image. But if we assume like in this communication that the downstream aberrations are fixed and known, this removes the degeneracy. 

\begin{acknowledgements}
This research is supported by the ``Groupement d'Int\'{e}r\^{e}t Scientifique'' PHASE.
\end{acknowledgements}

\bibliographystyle{aa}
\bibliography{mygouf}

\end{document}